\documentclass[10pt,twocolumn,letterpaper]{article}

\usepackage{cvpr}              
\usepackage[dvipsnames]{xcolor}
\usepackage{multirow}

\definecolor{cvprblue}{rgb}{0.21,0.49,0.74}
\usepackage[pagebackref,breaklinks,colorlinks,citecolor=cvprblue]{hyperref}
\usepackage[numbers]{natbib}

\newcommand{\OURS}{GRMM\xspace}

\title{GRMM: Real-Time High-Fidelity Gaussian Morphable Head Model \\with Learned Residuals}

\author{Mohit Mendiratta$^1$\thanks{Corresponding author: \texttt{mmendira@mpi-inf.mpg.de}}\quad
Mayur Deshmukh$^1$\quad
Kartik Teotia$^1$\quad
Vladislav Golyanik$^1$\quad
Adam Kortylewski$^1$$^,$$^2$\quad \\
Christian Theobalt$^1$\\
$^1$ Max Planck Institute for Informatics and Saarland University \\
$^2$ University of Freiburg\\
}

\begin{document}
\newcommand{\sysname}{GRMM\xspace}
\newcommand{\direction}{\mathbf{d}}

\newcommand{\loss}{\mathcal{L}}
\newcommand{\totalloss}{\loss_{total}}
\newcommand{\dataloss}{\loss_{img}}
\newcommand{\lightloss}{\loss_{lig}}
\newcommand{\regularizationloss}{\loss_{reg}}
\newcommand{\expressionloss}{\loss_{exp}}

\newcommand{\encoder}{\mathcal{E}}
\newcommand{\motionencoder}{\encoder_{m}}
\newcommand{\expressionencoder}{\encoder_{e}}
\newcommand{\transformencoder}{\encoder_{T}}
\newcommand{\decoder}{\Phi}
\newcommand{\meshdecoder}{\decoder_{mesh}}
\newcommand{\transformdecoder}{\decoder_{T}}
\newcommand{\opacitydecoder}{\decoder_{\alpha}}
\newcommand{\colordecoder}{\decoder_{app}}
\newcommand{\GHdecoder}{\decoder_{GH}}
\newcommand{\identitydecoder}{\decoder_{id}}

\newcommand{\latentcode}{z}
\newcommand{\expressioncode}{\latentcode_{e}}
\newcommand{\identitycode}{\latentcode_{id}}
\newcommand{\lightcode}{\latentcode_{l}}
\newcommand{\finalcode}{\latentcode_{f}}

\newcommand{\vertex}{\mathbf{v}_{\delta}}
\newcommand{\pixel}{p}
\newcommand{\blendweight}{\mathbf{w}}

\newcommand{\pos}{\mathbf{v}}
\newcommand{\templatepos}{\hat{\pos}}
\newcommand{\residualpos}{\delta\pos}
\newcommand{\featuremap}{\mathcal{F}}
\newcommand{\linearfeaturemap}{\featuremap_\mathrm{{lin}}}
\newcommand{\nonlinearfeaturemap}{\featuremap_\mathrm{{nlin}}}

\newcommand{\appearancefeaturemap}{\featuremap_\mathrm{{appe}}}
\newcommand{\alphafeaturemap}{\featuremap_\mathrm{{alpha}}}

\newcommand{\laplacian}{\mathbf{L}}
\newcommand{\blendshapes}{\mathbf{B}}

\newcommand{\voxelcolor}{c_{i}}
\newcommand{\voxelopacity}{o_{i}}

\newcommand{\opacityvalue}{I_{\alpha}}
\newcommand{\finalimage}{\hat{I}}
\newcommand{\renderedimage}{I_{rgb}}
\newcommand{\foregroundimage}{I_{rgb}}
\newcommand{\backgroundimage}{I_{BG}}

\newcommand{\cameraparams}{\phi}
\newcommand{\projection}{\Pi}

\twocolumn[{%
\renewcommand\twocolumn[1][]{#1}%
\maketitle
\vspace{-10pt}
\includegraphics[trim={0 0px 0 0},clip,width=\linewidth, height=5cm]{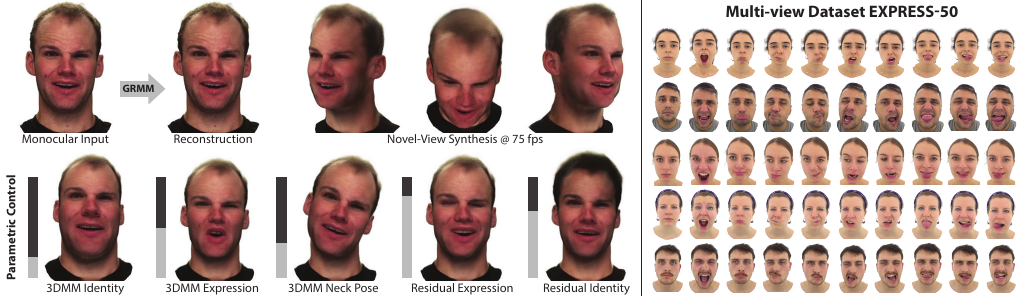}
\vspace{-20pt}
\captionof{figure}{  
\sysname provides disentangled control over a base 3DMM and learned residuals, fitting unseen identities from input images and enabling novel view synthesis and expression editing while preserving identity.  
It produces photorealistic 1K-resolution full-head renderings with diverse expressions in real time, achieving up to 75 fps.  
As part of this work, we also contribute EXPRESS-50, a dataset of 50 identities with 60 distinct expressions aligned across subjects, enabling consistent modeling of expression residuals.
}
\label{fig:teaser}

}]

\begin{abstract}
3D Morphable Models (3DMMs) enable controllable facial geometry and expression editing for reconstruction, animation, and AR/VR, but traditional PCA-based mesh models are limited in resolution, detail, and photorealism. Neural volumetric methods improve realism but remain too slow for interactive use.
Recent Gaussian Splatting (3DGS) based facial models achieve fast, high-quality rendering but still depend solely on a mesh-based 3DMM prior for expression control, limiting their ability to capture fine-grained geometry, expressions, and full-head coverage.
We introduce \sysname, the first full-head Gaussian 3D morphable model that augments a base 3DMM with residual geometry and appearance components, additive refinements that recover high-frequency details such as wrinkles, fine skin texture, and hairline variations. \sysname provides disentangled control through low-dimensional, interpretable parameters (e.g., identity shape, facial expressions) while separately modelling residuals that capture subject- and expression-specific detail beyond the base model’s capacity. Coarse decoders produce vertex-level mesh deformations, fine decoders represent per-Gaussian appearance, and a lightweight CNN refines rasterised images for enhanced realism, all while maintaining 75 FPS real-time rendering.
To learn consistent, high-fidelity residuals, we present EXPRESS-50, the first dataset with 60 aligned expressions across 50 identities, enabling robust disentanglement of identity and expression in Gaussian-based 3DMMs.
Across monocular 3D face reconstruction, novel-view synthesis, and expression transfer, \sysname surpasses state-of-the-art methods in fidelity and expression accuracy while delivering interactive real-time performance.
\end{abstract}
\vspace{-0.5em}    
\section{Introduction}
\label{sec:intro}
High-fidelity 3D face modelling is essential for VR, AR, animation, and digital avatar creation. A widely used class of methods, 3D Morphable Models (3DMMs) \cite{blanz1999morphable,egger20203d} compactly and controllably represent facial geometry and appearance using low-dimensional parametric spaces. However, achieving photorealism, real-time efficiency, and expressiveness remains challenging, as current methods struggle to capture fine details while maintaining real-time performance. 
Traditional mesh-based 3DMMs~\cite{blanz1999morphable,booth20163d,li2017learning,yang2020facescape} are efficient and interpretable but limited in resolution and unable to represent fine-scale geometry and texture variation. 
Neural rendering~\cite{yenamandra2020i3dmm,giebenhain2023nphm} and volumetric methods~\cite{hong2022headnerf,zhuang2022mofanerf,yang2024vrmm} improve visual fidelity, but are computationally heavy and struggle with large deformations or extreme expressions. More recently, 3D Gaussian Splatting (3DGS)\cite{kerbl3Dgaussians} has enabled high-resolution rendering at interactive rates. However, 3DGS-based facial models\cite{xu2024gphm,zheng2024headgap} still rely solely on coarse mesh-based 3DMM priors for expression control, limiting their \textit{expressivity} and ability to capture subtle, identity- or expression-specific detail. Furthermore, these models are not publicly available.

An additional bottleneck is the scarcity of datasets with both \textit{high expression diversity} and \textit{cross-identity expression alignment}. While recent datasets~\cite{pan2024renderme,kirschstein2023nersemble,martinez2024codec} improve identity coverage, they typically provide limited expression variability and lack the alignment required for disentangled identity–expression learning.

To address these limitations, we propose the Gaussian Residual Morphable Model (\OURS), the first open-source, full-head 3D Gaussian morphable model with learned residuals: additive refinements to both geometry and appearance that recover high-frequency details such as wrinkles, skin microstructure, and hairline variation beyond the capacity of a base 3DMM. GRMM offers independent control over (i) interpretable low-dimensional 3DMM parameters (identity shape, facial expression, head pose) and (ii) residual parameters that encode fine-scale, identity- and expression-specific deviations. This separation enables precise, compositional manipulation of facial attributes without sacrificing realism (Figure~\ref{fig:teaser}).

Technically, GRMM is based on a mesh-based base 3DMM~\cite{li2017learning} to produce a coarse head shape and predicts residual deformations using lightweight MLPs driven by low-dimensional identity and expression codes. UV-anchored Gaussian primitives deform coherently with the mesh, preserving spatial consistency and cross-identity correspondence. Convolutional decoders predict per-Gaussian appearance for detailed geometry and texture, while an image-space CNN refines rasterised images to recover surface details not fully captured by the Gaussians. The result is real-time 1K-resolution rendering at 75 FPS.

To support learning disentangled residuals, we introduce EXPRESS-50, a multi-view dataset containing 50 identities each performing 60 \textit{consistently aligned} expressions. Expression alignment is achieved via over 150 hours of frame-by-frame manual annotation, ensuring that all subjects exhibit semantically matched expressions (including challenging motions such as tongue movement). This alignment enables cross-identity supervision for robust residual disentanglement and improves generalisation to unseen subjects and expressions.
 
In summary, our contributions are:
 \begin{itemize}
 \item \OURS, the first open-source, full-head residual Gaussian morphable model that achieves high expressivity, fine-grained control, and real-time 1K rendering at 75 FPS, outperforming prior morphable face models in both quality and flexibility.
 \item A novel architecture that separates base 3DMM control from learned residual geometry and appearance, combined with enhanced mesh topology and UV mapping that explicitly models teeth and inner-mouth regions. Thus, enabling high rendering quality, speed, and expressivity.
 \item EXPRESS-50, i.e.,~a new multi-view image dataset with $50$ identities and $60$ aligned facial expressions, extends the corpus of existing datasets in the literature and serves as an essential ingredient to obtaining the results demonstrated in this paper.
 \end{itemize}
\section{Related Work}\label{sec:Related} 

\paragraph{Mesh-based head models.} Parametric 3D face models represent facial geometry, expression, and identity using low-dimensional parameters.
The seminal 3D Morphable Model (3DMM) by Blanz and Vetter~\cite{blanz1999morphable} aligns a fixed-topology mesh to 3D scans through non-rigid registration and learns shape and appearance spaces via PCA~\cite{abdi2010principal}.
Subsequent works~\cite{vlasic2005face, li2017learning} introduced multilinear models with separate control over facial components (e.g., jaw and eyes), which have become standard priors for reconstruction and tracking~\cite{thies2016face2face, Tewari_2017_ICCV, 10.1145/3197517.3201283}. 
FaceScape~\cite{yang2020facescape} improved realism with high-resolution geometry and diverse expressions, but mesh-based 3DMMs remain constrained by linear subspaces and limited expressiveness for fine details.
To overcome this, non-linear mesh 3DMMs~\cite{ranjan2018generating3dfacesusing, tran2018nonlinear3dfacemorphable, tran2019highfidelitynonlinear3dface} use deep networks to learn complex mappings from latent codes to mesh geometry, improving reconstruction quality and facial variation. However, these models often act as black boxes, sacrificing interpretability and editability, and typically remain limited to facial regions without supporting full-head modelling.
Delta models such as DECA~\cite{feng2021learning} and EMOCA~\cite{danvevcek2022emoca} enhance detail with UV-space displacements but remain restricted to the facial region.
Generative approaches, such as Morphable Diffusion~\cite{chen2024morphable}, leverage diffusion models conditioned on 3DMMs to synthesise avatars from a single image; however, they lack explicit control over identity and expression, and cannot represent the mouth interior or hair.
Our method unifies the controllability of mesh-based 3DMMs with learned full-head per-vertex deltas and 3D Gaussian refinement, thereby retaining interpretability while capturing high-frequency details that exceed the limits of linear or face-only models.
\paragraph{Implicit parametric head models.}  
Implicit representations have driven significant progress in neural parametric head modelling.
SDF-based methods~\cite{giebenhain2023nphm, yenamandra2020i3dmm} avoid fixed mesh topology and better capture complex structures like hair.
NeRF-based approaches~\cite{hong2022headnerf, wang2022morf, zhuang2022mofanerf} achieve photorealistic heads without explicit geometry, while hybrid techniques~\cite{lombardi2021mixture, cao2022authentic, yang2024vrmm} combine mesh priors with volumetric fields for improved controllability and realism, often using large-scale capture datasets.
However, NeRF-based models suffer from low rendering efficiency, forcing trade-offs between quality and speed. 
In contrast, our method predicts mesh-based deformations in a delta space and adds fine-scale details using 3D Gaussians, enabling efficient rendering while preserving control and high-frequency detail.
\paragraph{3D Gaussians-based head representations.}  
3D Gaussian splatting (3DGS) has recently emerged as a powerful approach for photorealistic novel-view rendering with real-time performance~\cite{kerbl3Dgaussians, wang2022voge}. 
Initially developed for rigid scenes, it has been extended to dynamic domains, including human heads and faces. 
The 3D Gaussian Parametric Head Model (GPHM)~\cite{xu2024gphm} adapts 3DGS for facial geometry by representing the head with a dense set of Gaussians trained on datasets such as NeRSemble~\cite{kirschstein2023nersemble} and FaceVerse~\cite{wang2022faceverse}, achieving high-quality synthesis. 
However, GPHM relies on MLP decoding, lacks a clear separation between coarse geometry and fine detail, and relies on 3DMM fitting and keypoints for reconstruction, which limits its expressiveness. 
HeadGAP~\cite{zheng2024headgap} builds on FLAME with part-based Gaussian offsets but remains constrained by FLAME’s fixed topology and shape space, while also inheriting the MLP overhead. 
Furthermore, HeadGAP cannot sample new identities or expressions, reducing its generative flexibility. 
Other re-enactment approaches, such as GAGAvatar~\cite{chu2024generalizable} and Portrait4D-v2~\cite{deng2024portrait4d}, use captured FLAME parameters but can only replay observed motions. 
Despite these advances, building a generative, expressive, and efficient head model remains an open challenge. 
Our approach combines mesh-based 3DMM control with full-head geometric deltas and 3D Gaussian refinement. We utilise a lightweight MLP for vertex geometry and convolutional decoders for per-Gaussian parameters, thereby avoiding the need for heavy per-Gaussian MLPs. This design reduces runtime, separates coarse geometry from fine details, and allows for sampling of identities and expressions.
\paragraph{Multiview head datasets.} Several multiview head datasets have been developed to advance 3D head modelling. FaceScape~\cite{yang2020facescape} captures 938 subjects with 20 expressions using high-resolution multiview images, primarily featuring East Asian identities and excluding hair. 
NeRSemble~\cite{kirschstein2023nersemble} comprises 300 identities in controlled setups, offering good subject diversity but limited expression coverage~\cite{prinzler2024jokerconditional3dhead}.
RenderMe-360~\cite{pan2024renderme} captures 500 subjects with full 360-degree views, including complex hairstyles and accessories, but offers only 12 expressions per subject. 
AVA-256~\cite{martinez2024codec} extends diversity by supporting 256 identities under consistent illumination and broad expression coverage, but it suffers from background matting and unnatural colour distribution, which complicates 3D reconstruction.
Although these datasets improve diversity and fidelity, they lack expression alignment across identities, which is critical for learning morphable models with precise identity-expression control.

We complement these datasets with our EXPRESS‑50 dataset, which provides expression alignment across 50 diverse identities. EXPRESS-50 captures a broader range of expressions than existing datasets, ensuring consistent correspondence across subjects. This alignment is essential for learning expressive, identity-disentangled morphable models.

\section{Method}
We present \sysname, a real-time, high-fidelity full-head morphable model that augments a mesh-based 3D Morphable Model (3DMM) with learned geometry and appearance residuals using 3D Gaussian splatting. 
Section~\ref{sec:datasets} introduces our new EXPRESS-50 dataset and the associated preprocessing pipeline, which together form a key contribution enabling consistent expression alignment across identities. 
Section~\ref{sec:Priliminaries} outlines the Gaussian attributes, image model, and camera model used to define the 3D representation and projection process. 
Section~\ref{sec:head_model_prior} describes the model structure, and Section~\ref{sec:training} details the training methodology. 
Finally, Section~\ref{sec:inference} presents the inference process, including refinement steps for full-head reconstruction.

\subsection{Expression-Aligned Datasets}
\label{sec:datasets}
We use two datasets to train our model: \textbf{EXPRESS-50}, a novel dataset we collected for complex facial expression modelling, and \textbf{RenderMe-360}~\cite{pan2024renderme}, a publicly available 4D human head dataset. EXPRESS-50 was created to provide a rich and diverse set of high-intensity expressions, complementing RenderMe-360’s broader identity coverage. EXPRESS-50 contains 50 identities performing 60 distinct expressions, recorded at 24 fps using a 24-camera FaceRig at 3840×2160 resolution. RenderMe-360 features 500 subjects performing 12 expressions, captured with 60 synchronised cameras at a resolution of 2448×2048.
\paragraph{Preprocessing.}
\label{alignment}
We preprocess both datasets through expression alignment, tracking, image preprocessing, and depth generation. A key contribution is the manual alignment of expressions across identities, enabling better disentanglement of identity and expression residuals in our morphable model. Age and gender statistics for EXPRESS-50 are in the supplemental (Section~\ref{sec:dataset_details}), and we will publicly release the dataset with annotations, expression labels, and preprocessing outputs.

\vspace{-1pt}
\paragraph{Expression Alignment.}
We manually annotate peak expressions in the EXPRESS-50 and RenderMe-360 datasets to ensure consistent expression alignment across identities.
In EXPRESS-50, each subject follows a scripted sequence of 60 expressions demonstrated via a reference video.
For a chosen reference identity, we manually select the frame where each expression is most prominent to serve as the canonical example.
For the remaining 49 identities, we select the peak expression frame that best matches each canonical frame, ensuring visual and semantic alignment across subjects.
In RenderMe-360, each of the 12 expressions is provided as a short video sequence per identity.
We observe that the final frame in each sequence typically captures the peak of the intended expression, so we annotate the last frame of each video as the aligned expression frame.
We use 280 RenderMe-360 identities for alignment, excluding those with heavy makeup or large accessories that interfere with facial appearance.
Together, these datasets comprise 330
unique identities, offering broad coverage of facial shapes and expressions.
Examples of consistent expression alignment are included in the supplementary material.

\paragraph{Tracking.}
To recover coarse facial geometry, we estimate FLAME~\cite{li2017learning} parameters: neck pose \( \boldsymbol{\theta}_{\text{neck}} \), jaw pose \( \boldsymbol{\theta}_{\text{jaw}} \), expression \( \boldsymbol{\alpha}_{\text{exp}} \), and identity \( \boldsymbol{\alpha}_{\text{id}} \). We adopt landmark- and photometry-based tracking using VHAP~\cite{qian2024vhap} to fit FLAME to annotated frames, obtaining a tracked mesh \( \mathbf{v}_{\text{rec}} \), global pose (\(\mathbf{R}, \mathbf{t}\)) per frame. For RenderMe-360 and EXPRESS-50, 68 facial landmarks from~\cite{bulat2017how} guide the tracking.
\paragraph{Ground-Truth Depth Generation.} 
To obtain high-quality ground truth depth images \( \mathbf{I}_{depth}^{gt} \) for supervision, we adopt ProbeSDF~\cite{toussaint2025probesdf}, a state-of-the-art surface reconstruction method. We apply it to each time step in our dataset to raycast depth from the optimised 3D surface.
\paragraph{Image Preprocessing.}
Foreground masks are extracted with RMBG-2.0~\cite{BiRefNet} for RenderMe-360 and BackgroundMattingV2~\cite{BGMv2} for EXPRESS-50, while Sapiens~\cite{khirodkar2024sapiens} provides additional facial masks to remove the torso and focus on the face.
\subsection{Preliminaries}
\label{sec:Priliminaries}

\paragraph{Image Generation Model.}
We build upon 3D Gaussian Splatting (3DGS)~\cite{kerbl3Dgaussians}, where each primitive is parameterized by position \(\mathbf{p}\), rotation \(\mathbf{r}\), scale \(\mathbf{s}\), opacity \(\mathbf{o}\), and color \(\mathbf{c}\). Following RaDe-GS~\cite{zhang2024rade}, we render depth from Gaussian attributes, and further attach a learnable feature vector \(\mathbf{f}_i \in \mathbb{R}^{32}\) to each primitive for richer mid-level appearance. The complete attribute set is
\begin{equation}
\label{eq:3dgs}
\mathcal{B} = \{ \mathbf{p}, \mathbf{r}, \mathbf{s}, \mathbf{o}, \mathbf{c}, \mathbf{f} \}, \quad 
\mathbf{I}_{rgb}, \mathbf{I}_{depth}, \mathbf{I}_{feature} = \mathcal{R}(\mathcal{B}, \pi_{\mathbf{K}, \mathbf{E}}),
\end{equation}
where \(\mathcal{R}\) is a differentiable rasterizer with camera intrinsics \(\mathbf{K}\) and extrinsics \(\mathbf{E}\). To ensure a consistent reference frame, we transform cameras into the canonical FLAME space:
\begin{equation}
\label{eq:rasterize}
\pi'_{\mathbf{R},\mathbf{t}} = \mathbf{K} \cdot \mathbf{E} \cdot \begin{bmatrix} \mathbf{R} & \mathbf{t} \end{bmatrix},
\end{equation}
where \(\mathbf{R},\mathbf{t}\) are the tracked FLAME mesh orientation and translation obtained from the preprocessing step Section~\ref{sec:datasets}.
\paragraph{GaussianHeads.}
Our \sysname method adapts the GaussianHeads (GH)~\cite{10.1145/3687927} architecture, which maps primitives to UV space of a template mesh. Unlike GH, designed for subject-specific reconstruction, our method generalizes across identities and expressions with a different formulation and training strategy. Given expression code \(\mathbf{z}_{exp}\) and view direction \(\direction\), GH predicts primitive attributes as
\begin{equation}
\{\mathbf{v}_\delta, \delta_r, \delta_p, \delta_s, \mathbf{o}, \mathbf{c}\} = \mathcal{D}_{GH}(\mathbf{z}_{exp}, \direction),
\end{equation}
where \(\mathbf{v}_\delta\) are mesh vertex deformations, \(\{\delta_r, \delta_p, \delta_s\}\) are rotation, translation and scale offsets relative to the template, and \(\mathbf{o}, \mathbf{c}\) denote opacity and color.  

\subsection{Gaussian Residual Morphable Model}
\label{sec:head_model_prior}
\begin{figure*}
    \centering    \includegraphics[width=\textwidth,height=6cm]{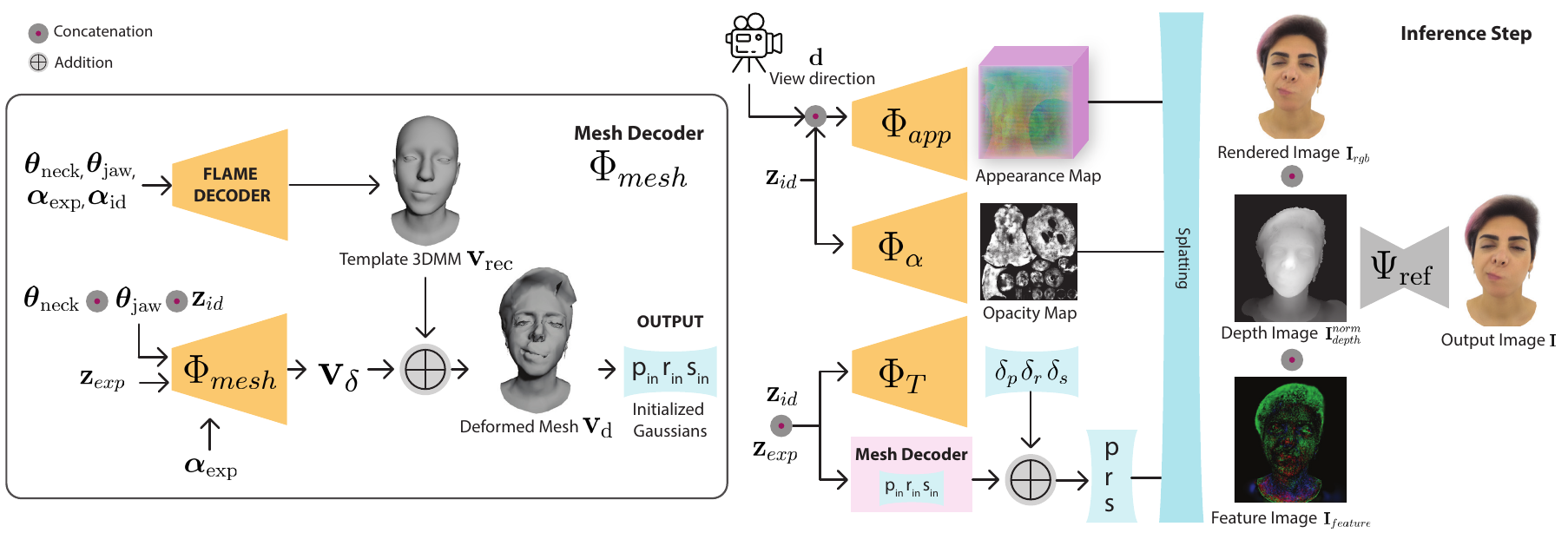}
    \caption{\textbf{Method pipeline.} 
    Identity and expression latents \(\mathbf{z}_{id}\in\mathbb{R}^{512}\) and \(\mathbf{z}_{exp}\in\mathbb{R}^{256}\), together with FLAME pose/expression parameters \((\theta_{\mathrm{neck}},\theta_{\mathrm{jaw}},\alpha_{\mathrm{exp}})\), drive the coarse mesh decoder \(\Phi_{\mathrm{mesh}}\) to predict per‑vertex displacements \(\mathbf{v}_{\delta}\). Adding these to the tracked mesh \(\mathbf{v}_{\mathrm{rec}}\) yields the deformed mesh \(\mathbf{M}_{\mathrm{d}}=(\mathbf{v}_{\mathrm{d}},\mathcal{F})\). UV‑anchored 3D Gaussians with initial \((\mathbf{p}_{\mathrm{in}},\mathbf{r}_{\mathrm{in}},\mathbf{s}_{\mathrm{in}})\) are placed on \(\mathbf{M}_{\mathrm{d}}\). The transformation decoder \(\Phi_{\mathrm{T}}(\mathbf{z}_{id},\mathbf{z}_{exp})\) outputs UV‑aligned maps \(\delta_p,\delta_r,\delta_s\) to refine position, rotation, and scale; the opacity decoder \(\Phi_{\alpha}(\mathbf{z}_{id})\) and appearance decoder \(\Phi_{\mathrm{app}} (\mathbf{z}_{id},\mathbf{d})\) produce opacity, RGB, and a 32‑D feature map. A differentiable rasterizer renders \(\mathbf{I}_{\mathrm{rgb}}, \mathbf{I}_{\mathrm{depth}}, \mathbf{I}_{\mathrm{feature}}\), where \(\mathbf{I}_{\mathrm{depth}}\) is normalized to \(\mathbf{I}_{depth}^{norm}\) and provided as input to the screen‑space CNN \(\Psi_{\mathrm{ref}}\), which outputs the final RGB image \(\mathbf{I}\).
    }
\label{fig:method_figure}
\end{figure*}

Our goal is to generate a high-fidelity head model for unseen identity and expression. Our method comprises a set of decoders (Figure~\ref{fig:method_figure}), $\mathcal{D}_{\text{\sysname}}:=\{ \meshdecoder, \transformdecoder, \opacitydecoder, \colordecoder \}$ and a refinement network $\Psi_\mathrm{ref}$. The mesh decoder $\meshdecoder$ predicts vertex deformations for \( \mathbf{v}_{\text{rec}} \), the transform decoder $\transformdecoder$ outputs Gaussian primitive transformations, the opacity decoder $\opacitydecoder$ estimates primitive opacities, and the color decoder $\colordecoder$ produces view-dependent appearance. All modules take as input the residual identity code $\mathbf{z}_{id}$, residual expression code $\mathbf{z}_{exp}$, neck pose \( \boldsymbol{\theta}_{\text{neck}} \), jaw pose \( \boldsymbol{\theta}_{\text{jaw}} \), expression coefficients \( \boldsymbol{\alpha}_{\text{exp}} \), and view direction $\direction$, generating a complete head model with photorealistic rendering.
\begin{equation}
\begin{aligned}
\{\vertex, \delta_r, \delta_p, \delta_s, \mathbf{o}, \mathbf{c}, \mathbf{f}\}
= \mathcal{D}_{\text{\sysname}}\!\big(
&\mathbf{z}_{id}, \mathbf{z}_{exp}, \boldsymbol{\alpha}_{\text{exp}}, \\
&\quad \boldsymbol{\theta}_{\text{neck}}, \boldsymbol{\theta}_{\text{jaw}}, \direction
\big).
\end{aligned}
\end{equation}
The refinement network $\Psi_\mathrm{ref}$ refines the rendered image $\mathbf{I_{rgb}}$, which is obtained from Equation \ref{eq:3dgs}. Each component is introduced in detail in the following sections.
\paragraph{Leveraging Expression Alignment.}
We represent each subject with a learnable residual identity latent code \( \mathbf{z}_{id} \in \mathbb{R}^{512} \) and each facial expression with a learnable global expression residual latent code \( \mathbf{z}_{exp} \in \mathbb{R}^{256} \). Each expression is associated with a single \( \mathbf{z}_{exp} \) that is shared across all identities. This design promotes a clear separation between identity and expression residuals.

\paragraph{Mesh Decoder}
The mesh decoder \( \meshdecoder \) predicts per-vertex identity and expression displacements. We use a shared MLP before the decoder to fuse identity and pose features, which improves conditioning and enables disentangled geometry prediction.  
A shared MLP processes the identity and pose inputs:
\begin{equation}
\mathbf{f}_{\text{base}} = \text{MLP}_{\text{shared}}([\mathbf{z}_{id}, \boldsymbol{\theta}_{\text{neck}}, \boldsymbol{\theta}_{\text{jaw}}]),
\end{equation}
where \(\mathbf{z}_{id} \in \mathbb{R}^{512}\) is the residual identity code, and \(\boldsymbol{\theta}_{\text{neck}}, \boldsymbol{\theta}_{\text{jaw}} \in \mathbb{R}^3\) are pose parameters.  
Identity displacements are predicted as: 
\begin{equation}
\mathbf{v}_{\delta,\text{id}} = \Phi_{\text{mesh,id}}(\mathbf{f}_{\text{base}}),
\end{equation}
where \(\mathbf{v}_{\delta,\text{id}} \in \mathbb{R}^{N_v \times 3}\) and \(N_v\) is the number of mesh vertices. 
FLAME expression modulation is applied to the concatenated identity–expression feature:
\begin{align}
\mathbf{f}_{\text{exp}} &= [\mathbf{f}_{\text{base}}, \mathbf{z}_{exp}], \quad \mathbf{z}_{exp} \in \mathbb{R}^{256}, \\
[\boldsymbol{\gamma}, \boldsymbol{\beta}] &= \text{MLP}_{\text{FiLM}}(\boldsymbol{\alpha}_{\text{exp}}), \quad \boldsymbol{\alpha}_{\text{exp}} \in \mathbb{R}^{100}, \\
\tilde{\mathbf{f}}_{\text{exp}} &= \mathbf{f}_{\text{exp}} + \boldsymbol{\gamma} \odot \mathbf{f}_{\text{exp}} + \boldsymbol{\beta},
\end{align}
where \(\text{MLP}_{\text{FiLM}}\) outputs the scale \(\boldsymbol{\gamma} \in \mathbb{R}^d\) and shift \(\boldsymbol{\beta} \in \mathbb{R}^d\) parameters for feature-wise linear modulation (FiLM).
Expression displacements are then computed as:
\begin{equation}
\mathbf{v}_{\delta,\text{exp}} =\Phi_{\text{mesh,exp}}
(\tilde{\mathbf{f}}_{\text{exp}}).
\end{equation}

The final deformed mesh is:
\begin{equation}
\mathbf{v}_{\mathrm{d}} = \mathbf{v}_{\mathrm{rec}} + \mathbf{v}_{\delta,\mathrm{id}} + \mathbf{M}_{\mathrm{face}} \, \mathbf{v}_{\delta,\mathrm{exp}},
\end{equation}
where \(\mathbf{M}_{\mathrm{face}}\) masks teeth vertices to prevent expression offsets. We add teeth vertices following VHAP and extend FLAME with inner-mouth faces, forming \(\mathbf{M}_{\mathrm{d}} = (\mathbf{v}_{\mathrm{d}}, \mathcal{F})\), where \(\mathcal{F}\) is the face connectivity. FLAME enhancements are provided and ablated in the supplemental material (Section ~\ref{sec:mesh_uv_enhancement}). Additionally, we ablate the importance of \( \meshdecoder \) in Section ~\ref{sec:additional_ablation}.
\vspace{-1em}
\paragraph{Gaussian Primitive Initialisation.}
We initialize \(\mathrm{N}_{\text{prim}} = \mathrm{N}_{\text{g}}^2\) 3D Gaussians by uniformly sampling the UV space of the 3DMM mesh at a resolution of \(\mathrm{N}_{\text{g}} \times \mathrm{N}_{\text{g}}\), with \(\mathrm{N}_{\text{g}} = 512\). Each Gaussian is assigned a position \(\mathbf{p}_{\text{in}} \in \mathbb{R}^3\) via barycentric interpolation on the deformed mesh \(\mathbf{M}_{\mathrm{d}}\) with vertices \(\mathbf{v}_{\mathrm{d}}\), and both rotation \(\mathbf{r}_{\text{in}} \in \mathbb{R}^3\) and scale \(\mathbf{s}_{\text{in}} \in \mathbb{R}^3\) are initialized to zero. To model the mouth interior, we enhance the FLAME UV map by defining separate UV regions for the teeth and mouth interior. More details are provided in the supplemental material (Section ~\ref{sec:mesh_uv_enhancement}).
\vspace{-1em}
\paragraph{Decoding the Gaussian Primitive Attributes.}
To enable high-resolution real-time rendering, we decode the properties of each Gaussian using efficient CNN decoders. Following~\cite{saito2024relightablegaussiancodecavatars, pang2024ashanimatablegaussiansplats}, these decoders map identity and expression codes \((\mathbf{z}_{id}, \mathbf{z}_{exp})\) to geometric and appearance attributes, capturing fine transformations and view-dependent colour.

\textbf{Transformation decoder} \(\transformdecoder\) maps \(\mathbf{z}_{id}\) and \(\mathbf{z}_{exp}\) to an offset map of size \(\mathrm{N}_{\text{g}} \times \mathrm{N}_{\text{g}} \times 10\), corresponding to the offsets of position (\(\delta_p\)), rotation (\(\delta_r\)) and scale (\(\delta_s\)) for the initial values \(\mathbf{p}_{\text{in}}\), \(\mathbf{r}_{\text{in}}\), and \(\mathbf{s}_{\text{in}}\). The updated Gaussian parameters are:
\begin{equation}
\mathbf{p} = \mathbf{p}_{\text{in}} + \delta_p, \quad 
\mathbf{r} = \delta_r, \quad 
\mathbf{s} = \delta_s.
\end{equation}
Position offsets \(\delta_p\) capture fine-scale surface variation, including facial hair and inner-mouth geometry. 

\textbf{Opacity decoder} \(\opacitydecoder\) predicts a map of size \(\mathrm{N}_{\text{g}} \times \mathrm{N}_{\text{g}} \times 1\), where each value represents the opacity \(\voxelopacity\) of a Gaussian and is only conditioned on \(\mathbf{z}_{id}\). 

\textbf{Appearance decoder} \(\colordecoder\) predicts a map of size \(\mathrm{N}_{\text{g}} \times \mathrm{N}_{\text{g}} \times 35\), where each entry contains RGB colour \(\voxelcolor \in \mathbb{R}^{3 \times 1}\) and a learned feature vector \(\mathbf{f}_i \in \mathbb{R}^{32 \times 1}\); this decoder is conditioned on \(\mathbf{z}_{id}\) and the view direction \(\direction\).
\vspace{-1em}
\paragraph{Refinement Network.}
We use a CNN in the screen space, $\Psi_{\mathrm{ref}}$, to refine the rendered results. The image resolution remains unchanged (1K) before and after refinement. Please refer to our video for a clearer illustration. We also ablate the importance of the refinement network in the supplementary material (Section~\ref{sec:additional_ablation}). This refinement enhances appearance priors that are difficult to capture for our 3DGS-based model without altering the underlying 3D representation, similar to the approaches in \cite{zheng2025headgapfewshot3dhead,xu2023latentavatar}.
\begin{align}
\label{eq:head-representation-cnn}
    &[\mathbf{I}_{rgb}, \mathbf{I}_{feature}, \mathbf{I}_{depth}] = \mathcal{R}(\mathcal{B}, \pi^{\prime}_{\mathbf{R},\mathbf{t}}),\\
    &\mathbf{I} = \Psi_{\mathrm{ref}}([\mathbf{I}_{rgb}, \mathbf{I}_{feature}, \mathbf{I}_{depth}^{norm}]),
\end{align}
The rendered depth image is standardised for \(\Psi_{\mathrm{ref}}\) by applying min--max normalisation, resulting in \(\mathbf{I}_{depth}^{norm}\).

\subsection{Training and Losses} 
\label{sec:training}
Given the \sysname representation, our proposed model is learned end-to-end using multi-view image supervision to train the decoders and refinement network.
For this, we optimise the following objective function:
\begin{align}
\mathbf{L}\ = & \ \mathbf{L}_{\text{rec}}(\mathbf{I}, \mathbf{I}^{*})\ + \mathbf{L}_{\text{rec}}(\mathbf{I}_{rgb}, \mathbf{I}^{*})\ + \notag \\
    & \lambda_{depth} \cdot \mathbf{L}_{\text{depth}}(\mathbf{I}_{depth}, \mathbf{I}_{depth}^{gt})\ + \mathbf{L}_{\text{reg}}\ .
\end{align}
Here, \( \mathbf{I}^{*} \) denotes the ground-truth RGB image. \( \mathbf{L}_{\text{rec}} \) is the reconstruction loss computed between both the image-space prediction \( \mathbf{I} \) from refinement network and the rendered image \( \mathbf{I}_{rgb} \) against \( \mathbf{I}^{*} \). \( \mathbf{L}_{\text{depth}} \) is the L2 loss between the predicted and ground-truth depth images, scaled by the weight \( \lambda_{depth} \). Finally, \( \mathbf{L}_{\text{reg}} \) represents additional regularization terms applied during training.
Specifically, image reconstruction loss:
\begin{equation}
\label{eq:reconstruction_loss}
\begin{array}{c}
    \mathbf{L}_{\text{rec}} = \lambda_{l1}\mathbf{L}_{\text{l1}} + \lambda_{ssim}\mathbf{L}_{\text{ssim}} + \lambda_{perc}\mathbf{L}_{\text{perc}}
\end{array}
\end{equation}
consists of L1 loss $\mathbf{L}_{\text{l1}}$, SSIM loss $\mathbf{L}_{\text{ssim}}$, and perceptual loss $\mathbf{L}_{\text{perc}}$ with the VGG \cite{simonyan2015deepconvolutionalnetworkslargescale} as the backbone.
Meanwhile, the training regularization loss:
\begin{equation}
\label{eq:regularization_loss}
\begin{array}{c}
    \mathbf{L}_{reg} = \lambda_{s}\mathbf{L}_{\text{s}} + \lambda_{z}\mathbf{L}_{\text{z}} + \lambda_{lap}\mathbf{L}_{\text{lap}},
\end{array}
\end{equation}
Here, \( \mathbf{L}_{\text{s}} \) is a regularisation term on the scale parameters, which encourages the scale of Gaussian primitives $\mathbf{s}$ to stay within a constrained range as follows:
\begin{equation}
\begin{aligned}
\label{eq:loss_reg2}
\mathbf{L}_{\mathrm{s}} = \mathrm{mean}(l_{\mathrm{s}}), \, l_{\mathrm{s}} &= 
\begin{cases}
1/\max(s_{i,d} ,10^{-7})  & \text{if } s_{i,d} < 0.1 \\
(s_{i,d} - 10.0)^2 & \text{if } s_{i,d} > 10.0\\
0 & \text{otherwise},
\end{cases}
\end{aligned}
\end{equation}
where $s_{i,d},d \in \{x, y, z\}$ denotes the scale value of each Gaussian primitive $\mathbf{i}$ along each axis, and $\mathrm{mean}(\cdot)$ is the average operation across all dimensions,
similar to \cite{saito2023relightable}.
\( \mathbf{L}_{\text{lap}} \) represents a smoothness regularization term for the deformed mesh \( \mathbf{M}_{d} \), and \( \mathbf{L}_{\text{z}} \) is the \( \mathbf{L}_{2} \)-norm of $\mathbf{z}_{id}$ and $\mathbf{z}_{exp}$ to improve the disentanglement.

 In our experiments, we set $\lambda_{l1}=0.8$, $\lambda_{ssim}=0.2$, $\lambda_{perc}=0.04$, $\lambda_z=0.001$, $\lambda_{lap}=0.01$, $\lambda_{s}=0.1$ and $\lambda_{depth}=0.5$. 
 
\subsection{Fitting via Inverse Rendering}
\label{sec:inference}
Given a single- or multi-view RGB portrait, we obtain 3D face tracking with VHAP and align inputs (Sec.~\ref{alignment}). We use a two-stage optimisation.

\textbf{Stage 1 (latent inversion).}
With decoders fixed, we optimise the latent codes \(\mathbf{z}_{id}\) and \(\mathbf{z}_{exp}\) by minimising
\begin{align}
\mathbf{L}^{(1)}_{\mathrm{fit}}
=~ \mathbf{L}_{\mathrm{rec}}(\mathbf{I}, \mathbf{I}^*)
+ \mathbf{L}_{\mathrm{rec}}(\mathbf{I}_{\mathrm{rgb}}, \mathbf{I}^*)
+ \lambda_{z}\,\mathbf{L}_{z}.
\end{align}

\textbf{Stage 2 (prior-preserving refinement).}
We then fix \(\mathbf{z}_{id}\) and \(\mathbf{z}_{exp}\) and fine-tune \(\mathcal{D}_{\text{\sysname}}\) for by minimising
\begin{align}
\mathbf{L}^{(2)}_{\mathrm{fit}}
=~ &\mathbf{L}_{\mathrm{rec}}(\mathbf{I}, \mathbf{I}^*)
+ \mathbf{L}_{\mathrm{rec}}(\mathbf{I}_{\mathrm{rgb}}, \mathbf{I}^*) + \lambda_{loc}\,\mathbf{L}_{\mathrm{loc}}.
\end{align}
\(\mathbf{L}_{\mathrm{loc}}\) is a PTI~\cite{roich2021pivotaltuninglatentbasedediting}-inspired locality regulariser that constrains updates to a small neighbourhood of the pretrained solution, preserving the prior. We set \(\lambda_{loc}=0.1\). We define and ablate \(\mathbf{L}_{\mathrm{loc}}\) in detail in the supplementary material (Section ~\ref{sec:locality}).
\section{Experiments}\label{sec:Experiments}
\label{sec:Experiments}
We evaluate \sysname on the RAVDESS~\cite{livingstone2018ryerson} dataset for monocular 3D face reconstruction, using 10 randomly selected identities. To assess novel-view synthesis across diverse expressions, we further sample ten identities from NeRSemble~\cite{kirschstein2023nersemble}, five from RenderMe, and three from EXPRESS-50. RAVDESS provides monocular RGB videos with acted emotions, while NeRSemble contributes multi-view recordings that capture complex expressions and head motion. We report ablation studies in Section~\ref{sec:Experiments:Evaluation}, and present results on downstream applications including monocular fitting, novel-view synthesis, and expression transfer in Section~\ref{sec:Experiments:Application}. Additionally, examples of disentangled parametric control for the inverted identities are included in the supplementary material (Section~\ref{sec:additional_results}).
\subsection{Ablation}
\label{sec:Experiments:Evaluation}
\begin{figure}[h!]
\centering
\includegraphics[width=\linewidth]{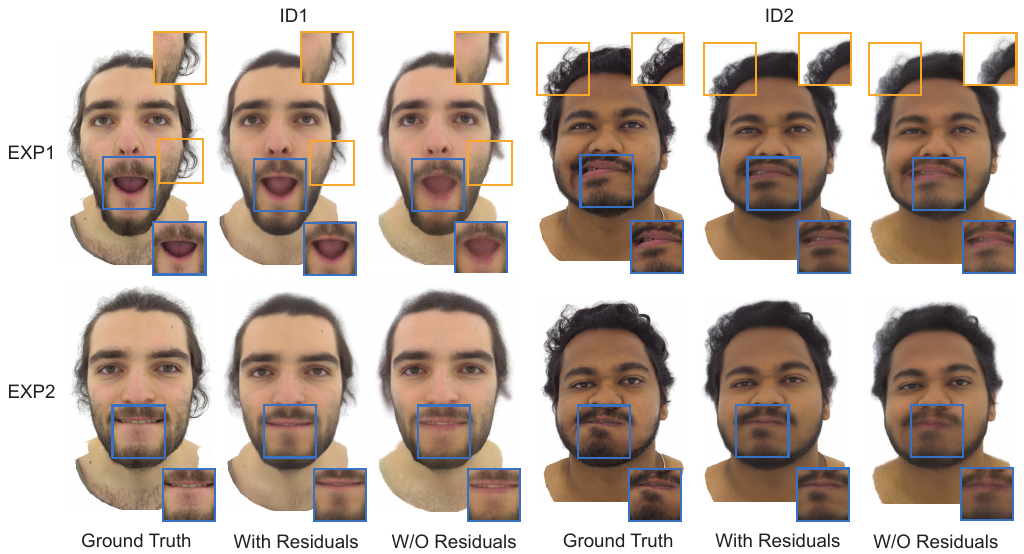}
\caption{\textbf{Residual parameterization improves fidelity.} Qualitative ablation comparing \emph{W/O residuals} vs. \emph{With residuals}. Residuals yield finer hair detail and better mouth articulation (e.g., for ID1, EXP2 the mouth cannot roll in without residuals), with higher PSNR (dB): \emph{W/O residuals} 28.91 vs. \emph{With residuals} 30.54 (\(+1.63\)). \textit{Please zoom in for details.}
}
\label{fig:ablation_residuals}
\end{figure}

\paragraph{No-residuals (direct 3DMM conditioning).} In this variant, we remove residual parameterisation and condition the network directly on the FLAME parameters.
\begin{equation}
\begin{aligned}
\{\vertex, \delta_r, \delta_p, \delta_s, \mathbf{o}, \mathbf{c}, \mathbf{f}\}
= \mathcal{D}_{\text{\sysname}}\!\big(
 \boldsymbol{\alpha}_{\text{id}},
 \boldsymbol{\alpha}_{\text{exp}},  \boldsymbol{\theta}_{\text{neck}}, \boldsymbol{\theta}_{\text{jaw}}, \direction
\big).
\end{aligned}
\end{equation}
Learning residuals \(\mathbf{z}_{id}\) and \(\mathbf{z}_{exp}\) over FLAME parameters enhances identity and expression representations, yielding finer hair and appearance details and improved mouth articulation (see Figure~\ref{fig:ablation_residuals}).

\begin{figure}[h!]
   \centering    \includegraphics[width=0.5\textwidth]{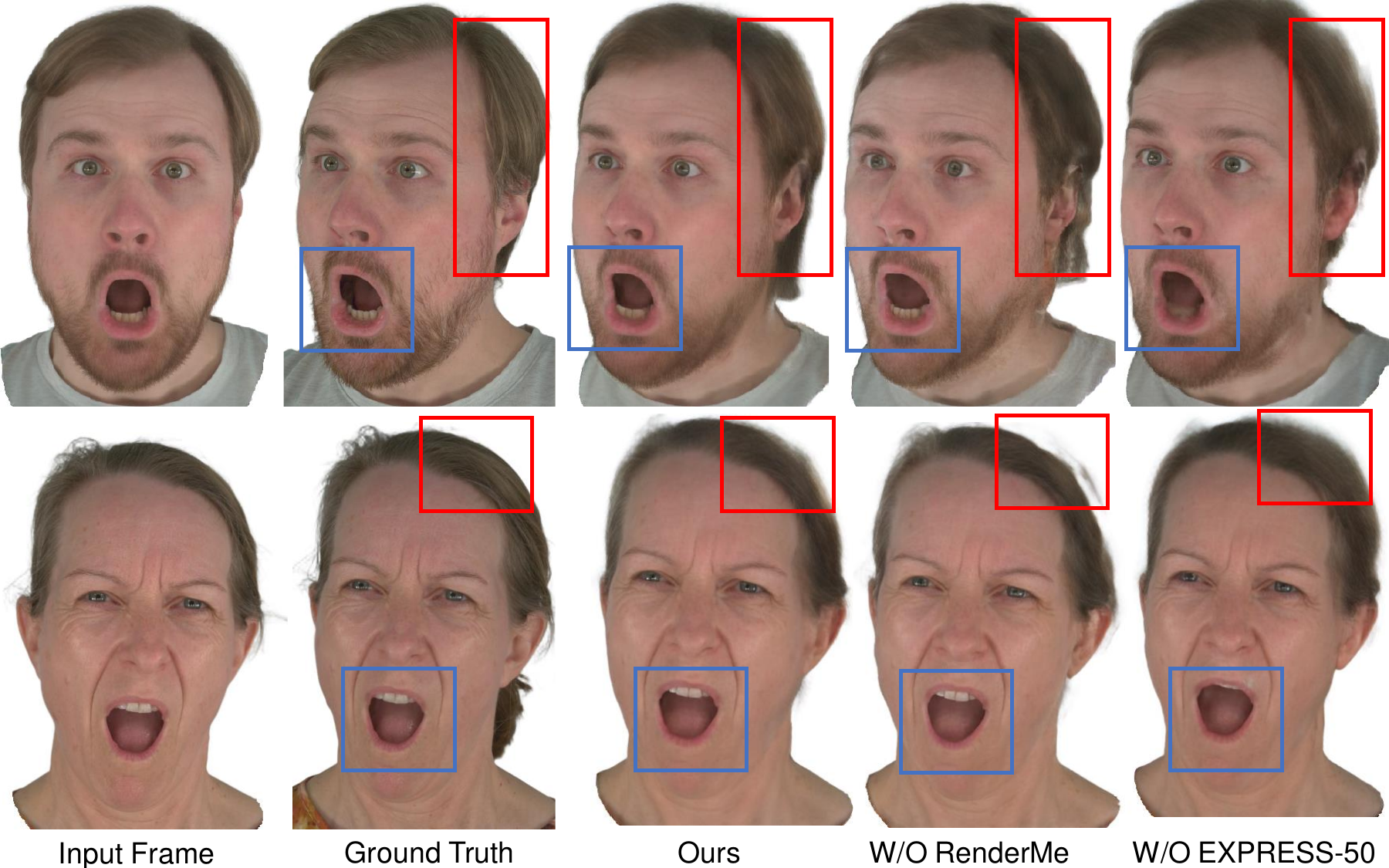} \caption{\textbf{Combining Datasets.} \textit{(left to right)} Ground Truth, Without EXPRESS-50, without RenderMe and  Ours. Our model trained with the combined datasets leads to overall better identity and expression fidelity. 
   } \label{fig:mouth_loss}
\end{figure}

\vspace{-2em}
\paragraph{Combining Datasets.}We conduct an ablation study to assess the impact of combining EXPRESS-50 and RenderMe. Using camera views as input, we fit our model to target identities from NeRSemble. Figure~\ref{fig:mouth_loss} and Table~\ref{tab:mouth_loss} compare models trained without EXPRESS-50, without RenderMe, and with both datasets. Joint training clearly improves identity and expression fidelity, highlighting the complementary strengths of RenderMe-360 for identity generalization and EXPRESS-50 for expression generalization.
\vspace{-0.5em}
\begin{table}[]
\centering
\caption{\textbf{Combining Datasets.} Joint training clearly enhances both identity and expression fidelity, showcasing the complementary strengths of RenderMe-360 for identity generalization and EXPRESS-50 for expression generalization.}
\begin{tabular}{l|ccc}
 
\hline
Method   & PSNR $\uparrow$         & SSIM $\uparrow$        & LPIPS $\downarrow$   \\ \hline
W/O EXPRESS-50 &  24.56         & 0.87        & 0.126        \\
W/O RenderMe & 25.27         & 0.90      & 0.115       \\
Ours - Full Model     & \textbf{27.40}         & \textbf{0.92}        & \textbf{0.091}  \\ \hline
\end{tabular}
\label{tab:mouth_loss}%
\end{table}
\subsection{Comparisons and Application}
\label{sec:Experiments:Application}
In this section, we demonstrate applications of \sysname in monocular image fitting, novel-view synthesis, and expression transfer. These applications showcase the generalisation capacity of our model to unseen identities, expressions and views. 
\vspace{-1em}
\paragraph{Compared Methods.}
We compare \sysname with publicly available parametric head models, including HeadNeRF~\cite{hong2022headnerf}, MoFaNeRF~\cite{zhuang2022mofanerf}, and the recent Morphable Diffusion~\cite{chen2024morphable}.
We observe that Morphable Diffusion is not a volume-rendering approach and achieves viewpoint control via a conditioning camera. Consequently, it does not generalise well to camera distributions from unseen datasets; when rendering novel viewpoints using evaluation dataset cameras, the results appear misaligned and slightly distorted. Therefore, we do not report quantitative metrics for Morphable Diffusion, but instead assess its performance through a user study and qualitative analysis.
The user study evaluates novel view, expression and identity consistency, with full details and results provided in the supplementary material (Section ~\ref{sec:user_study}).
\begin{figure*}
    \centering    \includegraphics[width=\textwidth]{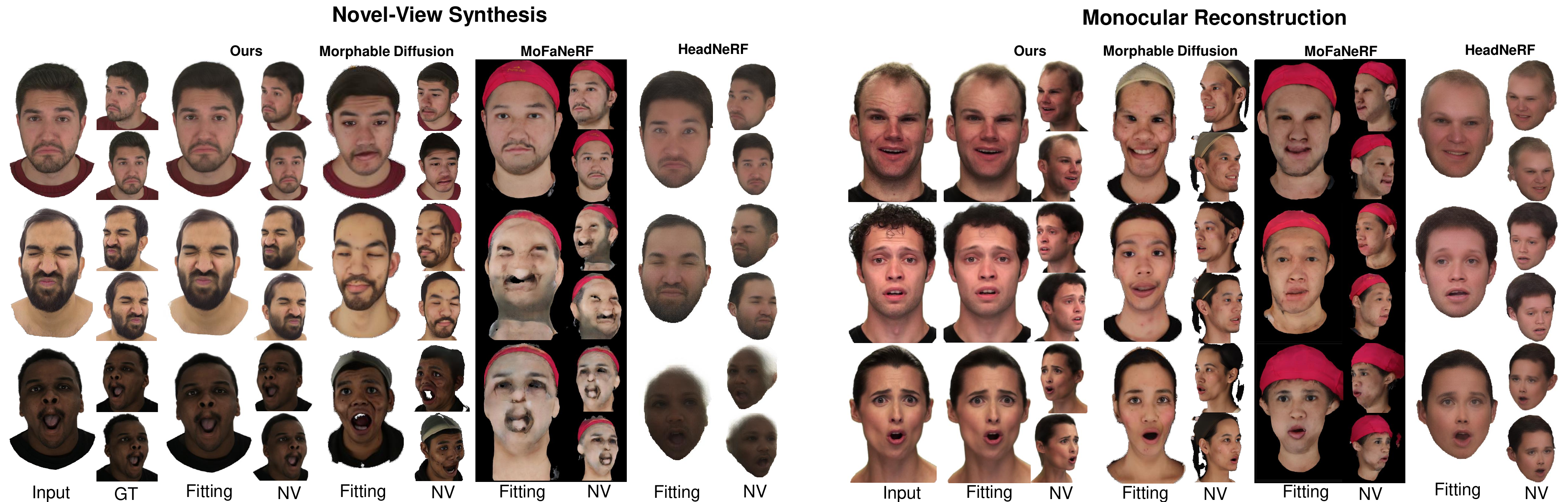}
    \caption{\textbf{Novel view synthesis (left) and monocular reconstruction (right).}   
    Left: given one or few posed views, \sysname synthesizes unseen viewpoints while preserving identity and expression.  
    Right: from a single RGB frame, \sysname reconstructs the subject and renders both the input and novel views.  
    We compare against Morphable Diffusion~\cite{chen2024morphable}, MoFaNeRF~\cite{zhuang2022mofanerf}, and HeadNeRF~\cite{hong2022headnerf} in both settings.  
    FID is reported only in the monocular inversion setting, where \sysname achieves lower FID than all other baselines.  
    \textit{Please zoom in for details.}}

    \label{fig:fitting}
\end{figure*}

\begin{table}[]
\centering
\caption{\textbf{Novel-view synthesis.} Quantitative comparison on held-out views. \sysname achieves the best performance, substantially improving over MoFaNeRF and HeadNeRF.}
\begin{tabular}{l|ccc}
 
\hline
Method   & PSNR $\uparrow$         & SSIM $\uparrow$        & LPIPS $\downarrow$   \\ \hline
MoFaNeRF  & 13.89         & 0.55        & 0.372       \\
HeadNeRF & 17.42        & 0.85       & 0.178       \\
Ours    & \textbf{30.85}         & \textbf{0.97}        & \textbf{0.072}  \\ \hline
\end{tabular}
\label{tab:novel_single_image}%
\end{table}

\begin{table}[]
\centering
\caption{\textbf{Monocular reconstruction.} Quantitative comparison using RMSE and FID. \sysname outperforms MoFaNeRF and HeadNeRF, indicating improved pixel accuracy and perceptual fidelity.
}

\begin{tabular}{l|ccc}
 
\hline
Method   & RMSE $\downarrow$         & FID $\downarrow$  \\ \hline
MoFaNeRF  & 0.193  & 290.786 \\
HeadNeRF & 0.067   & 116.34  \\
Ours    & \textbf{0.022}         & \textbf{74.34} \\ \hline
\end{tabular}
\label{tab:single_image}%
\end{table}
\vspace{-1em}
\paragraph{Novel-view Synthesis.} We evaluate \sysname for the task of novel-view synthesis for different identities in our evaluation set. We fit our model to a single viewpoint as described in Section \ref{sec:inference}, and assess our model's performance on two holdout views. Our method shows improved fitting and novel-view synthesis quality, as shown in Figure~\ref{fig:fitting} and Table~\ref{tab:novel_single_image}. Note that for the related methods, we use their publicly available inference code without any modifications.
\begin{figure}[h!]
    \includegraphics[width=\linewidth]{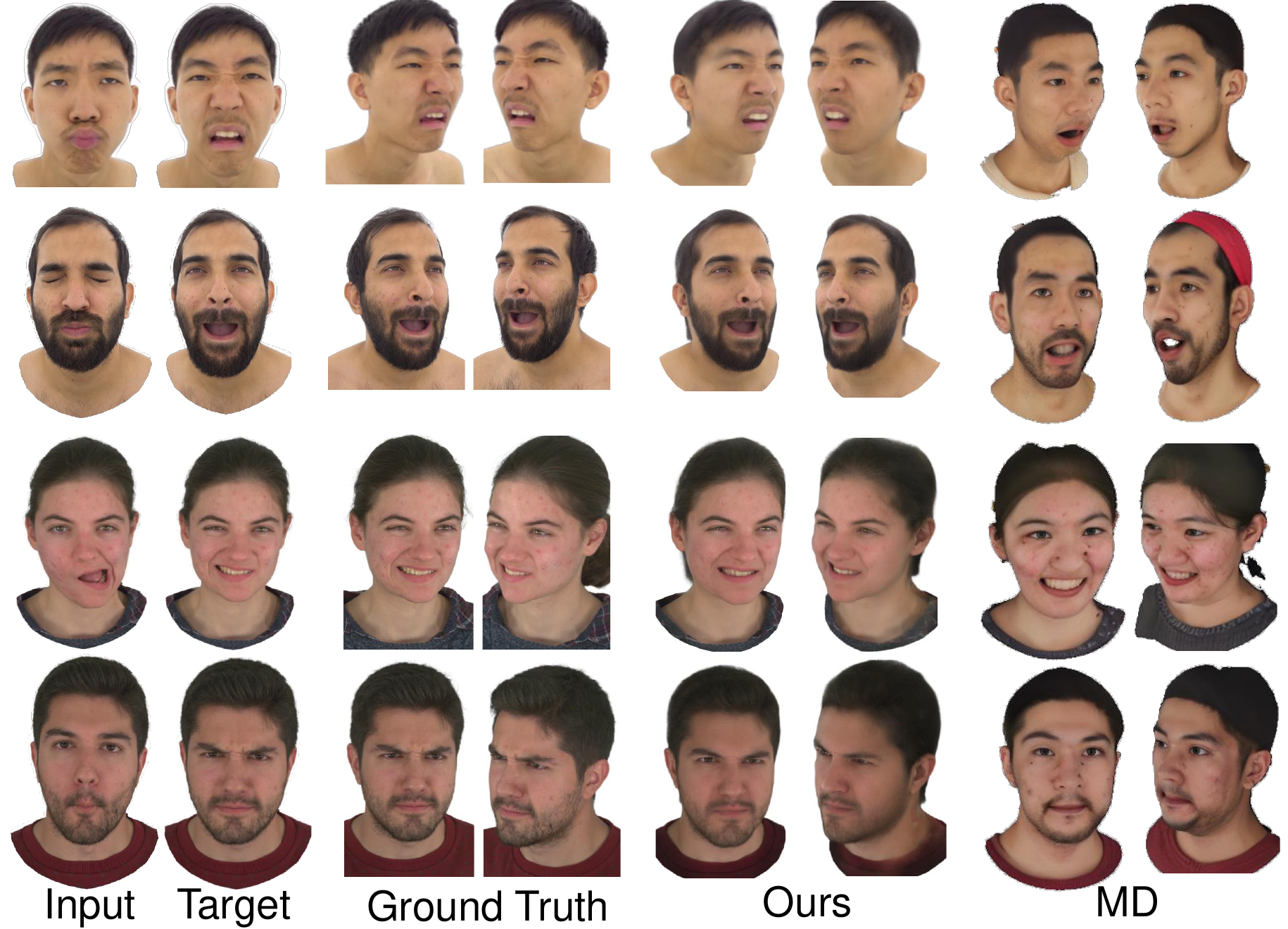}
    \caption{\textbf{Expression transfer.} From a single frontal image, we invert (Sec.~\ref{sec:inference}), swap expression parameters, and render novel views on EXPRESS-50 and NeRSemble. \sysname preserves identity and subtle expressions with multi-view consistency, whereas Morphable Diffusion (MD) generates inconsistent expressions across views. \textit{Please zoom in for details.}}
    \label{fig:expr_transfer}
\end{figure}
\vspace{-1em}
\paragraph{Expression Transfer.}
We compare \sysname to Morphable Diffusion for expression transfer by randomly sampling target expressions for selected identities from EXPRESS-50 and NeRSemble. 
For each identity, we perform inversion (Section ~\ref{sec:inference}) on a single frontal view, swap the expression parameters, and render the results under novel views. 
Qualitatively, Morphable Diffusion struggles to capture subtle expressions and produces expression-inconsistent renderings across novel views, whereas \sysname preserves both identity and expressions with multi-view consistency at high resolution and real-time frame rates (see Figure~\ref{fig:expr_transfer} and the supplemental video). 
A user study (Section ~\ref{sec:user_study}) corroborates these findings: participants consistently preferred \sysname for both expression accuracy and identity preservation in side-by-side novel-view comparisons.
\vspace{-1em}
\subsection{Conclusion}
We present \sysname, a Gaussian Residual Morphable Model that addresses key limitations of existing 3D morphable head models. By combining the efficiency of 3D Gaussian Splatting with residual augmentation of a mesh-based 3DMM in a coarse-to-fine pipeline, GRMM achieves photorealistic and diverse facial expressions at 1K resolution in real time (75 fps). The model leverages UV-anchored Gaussians and a lightweight refinement stage to enhance rendering quality, speed, and controllable expressivity. We also introduce EXPRESS-50, a multi-view dataset with 50 identities and 60 semantically aligned expressions, and show that joint training with RenderMe-360 improves both identity generalisation and expression fidelity. While advancing the state of the art, GRMM remains challenged by out-of-distribution subjects and lighting variations, suggesting the need for more diverse training data to improve robustness. With its ability to capture exaggerated expressions and render interactively, GRMM is well-suited for applications in computer graphics, virtual and augmented reality, and facial animation.
\vspace{-1.5em}
{
    \small
    \bibliographystyle{ieeenat_fullname}
    \bibliography{main}

\begin{thebibliography}{52}
\providecommand{\natexlab}[1]{#1}
\providecommand{\url}[1]{\texttt{#1}}
\expandafter\ifx\csname urlstyle\endcsname\relax
  \providecommand{\doi}[1]{doi: #1}\else
  \providecommand{\doi}{doi: \begingroup \urlstyle{rm}\Url}\fi

\bibitem[Abdi and Williams(2010)]{abdi2010principal}
Herv{\'e} Abdi and Lynne~J Williams.
\newblock Principal component analysis.
\newblock \emph{Wiley interdisciplinary reviews: computational statistics}, 2\penalty0 (4):\penalty0 433--459, 2010.

\bibitem[Blanz and Vetter(1999)]{blanz1999morphable}
V Blanz and T Vetter.
\newblock A morphable model for the synthesis of 3d faces.
\newblock In \emph{26th Annual Conference on Computer Graphics and Interactive Techniques (SIGGRAPH 1999)}, pages 187--194. ACM Press, 1999.

\bibitem[Booth et~al.(2016)Booth, Roussos, Zafeiriou, Ponniah, and Dunaway]{booth20163d}
James Booth, Anastasios Roussos, Stefanos Zafeiriou, Allan Ponniah, and David Dunaway.
\newblock A 3d morphable model learnt from 10,000 faces.
\newblock In \emph{Proceedings of the IEEE conference on computer vision and pattern recognition}, pages 5543--5552, 2016.

\bibitem[Bulat and Tzimiropoulos(2017)]{bulat2017how}
Adrian Bulat and Georgios Tzimiropoulos.
\newblock How far are we from solving the 2d \& 3d face alignment problem? (and a dataset of 230,000 3d facial landmarks).
\newblock In \emph{International Conference on Computer Vision}, 2017.

\bibitem[Cao et~al.(2022)Cao, Simon, Kim, Schwartz, Zollhoefer, Saito, Lombardi, Wei, Belko, Yu, Sheikh, and Saragih]{cao2022authentic}
Chen Cao, Tomas Simon, Jin~Kyu Kim, Gabe Schwartz, Michael Zollhoefer, Shun-Suke Saito, Stephen Lombardi, Shih-En Wei, Danielle Belko, Shoou-I Yu, Yaser Sheikh, and Jason Saragih.
\newblock Authentic volumetric avatars from a phone scan.
\newblock \emph{ACM Trans. Graph.}, 41\penalty0 (4), 2022.

\bibitem[Chen et~al.(2024)Chen, Mihajlovic, Wang, Prokudin, and Tang]{chen2024morphable}
Xiyi Chen, Marko Mihajlovic, Shaofei Wang, Sergey Prokudin, and Siyu Tang.
\newblock Morphable diffusion: 3d-consistent diffusion for single-image avatar creation.
\newblock In \emph{Proceedings of the IEEE/CVF Conference on Computer Vision and Pattern Recognition}, pages 10359--10370, 2024.

\bibitem[Chu and Harada(2024)]{chu2024generalizable}
Xuangeng Chu and Tatsuya Harada.
\newblock Generalizable and animatable gaussian head avatar.
\newblock \emph{Advances in Neural Information Processing Systems}, 37:\penalty0 57642--57670, 2024.

\bibitem[Dan{\v{e}}{\v{c}}ek et~al.(2022)Dan{\v{e}}{\v{c}}ek, Black, and Bolkart]{danvevcek2022emoca}
Radek Dan{\v{e}}{\v{c}}ek, Michael~J Black, and Timo Bolkart.
\newblock Emoca: Emotion driven monocular face capture and animation.
\newblock In \emph{Proceedings of the IEEE/CVF Conference on Computer Vision and Pattern Recognition}, pages 20311--20322, 2022.

\bibitem[Deng et~al.(2024)Deng, Wang, and Wang]{deng2024portrait4d}
Yu Deng, Duomin Wang, and Baoyuan Wang.
\newblock Portrait4d-v2: Pseudo multi-view data creates better 4d head synthesizer.
\newblock In \emph{European Conference on Computer Vision}, pages 316--333. Springer, 2024.

\bibitem[Egger et~al.(2020)Egger, Smith, Tewari, Wuhrer, Zollhoefer, Beeler, Bernard, Bolkart, Kortylewski, Romdhani, et~al.]{egger20203d}
Bernhard Egger, William~AP Smith, Ayush Tewari, Stefanie Wuhrer, Michael Zollhoefer, Thabo Beeler, Florian Bernard, Timo Bolkart, Adam Kortylewski, Sami Romdhani, et~al.
\newblock 3d morphable face models—past, present, and future.
\newblock \emph{ACM Transactions on Graphics (ToG)}, 39\penalty0 (5):\penalty0 1--38, 2020.

\bibitem[Feng et~al.(2021)Feng, Feng, Black, and Bolkart]{feng2021learning}
Yao Feng, Haiwen Feng, Michael~J Black, and Timo Bolkart.
\newblock Learning an animatable detailed 3d face model from in-the-wild images.
\newblock \emph{ACM Transactions on Graphics (ToG)}, 40\penalty0 (4):\penalty0 1--13, 2021.

\bibitem[Giebenhain et~al.(2023)Giebenhain, Kirschstein, Georgopoulos, R{\"{u}}nz, Agapito, and Nie{\ss}ner]{giebenhain2023nphm}
Simon Giebenhain, Tobias Kirschstein, Markos Georgopoulos, Martin R{\"{u}}nz, Lourdes Agapito, and Matthias Nie{\ss}ner.
\newblock Learning neural parametric head models.
\newblock In \emph{Proc. IEEE Conf. on Computer Vision and Pattern Recognition (CVPR)}, 2023.

\bibitem[Haotian et~al.(2024)Haotian, Mingwu, ChongYang, Yu-Kun, Pengfei, and Haibin]{yang2024vrmm}
Yang Haotian, Zheng Mingwu, Ma ChongYang, Lai Yu-Kun, Wan Pengfei, and Huang Haibin.
\newblock Vrmm: A volumetric relightable morphable head model.
\newblock In \emph{SIGGRAPH 2024 Conference Proceedings}, 2024.

\bibitem[Hong et~al.(2022)Hong, Peng, Xiao, Liu, and Zhang]{hong2022headnerf}
Yang Hong, Bo Peng, Haiyao Xiao, Ligang Liu, and Juyong Zhang.
\newblock Headnerf: A real-time nerf-based parametric head model.
\newblock In \emph{Proceedings of the IEEE/CVF Conference on Computer Vision and Pattern Recognition}, pages 20374--20384, 2022.

\bibitem[Kerbl et~al.(2023)Kerbl, Kopanas, Leimk{\"u}hler, and Drettakis]{kerbl3Dgaussians}
Bernhard Kerbl, Georgios Kopanas, Thomas Leimk{\"u}hler, and George Drettakis.
\newblock 3d gaussian splatting for real-time radiance field rendering.
\newblock \emph{ACM Transactions on Graphics}, 42\penalty0 (4), 2023.

\bibitem[Khirodkar et~al.(2024)Khirodkar, Bagautdinov, Martinez, Zhaoen, James, Selednik, Anderson, and Saito]{khirodkar2024sapiens}
Rawal Khirodkar, Timur Bagautdinov, Julieta Martinez, Su Zhaoen, Austin James, Peter Selednik, Stuart Anderson, and Shunsuke Saito.
\newblock Sapiens: Foundation for human vision models.
\newblock \emph{arXiv preprint arXiv:2408.12569}, 2024.

\bibitem[Kim et~al.(2018)Kim, Garrido, Tewari, Xu, Thies, Niessner, P\'{e}rez, Richardt, Zollh\"{o}fer, and Theobalt]{10.1145/3197517.3201283}
Hyeongwoo Kim, Pablo Garrido, Ayush Tewari, Weipeng Xu, Justus Thies, Matthias Niessner, Patrick P\'{e}rez, Christian Richardt, Michael Zollh\"{o}fer, and Christian Theobalt.
\newblock Deep video portraits.
\newblock \emph{ACM Trans. Graph.}, 37\penalty0 (4), 2018.

\bibitem[Kingma and Ba(2017)]{kingma2017adammethodstochasticoptimization}
Diederik~P. Kingma and Jimmy Ba.
\newblock Adam: A method for stochastic optimization, 2017.

\bibitem[Kirschstein et~al.(2023)Kirschstein, Qian, Giebenhain, Walter, and Nie\ss{}ner]{kirschstein2023nersemble}
Tobias Kirschstein, Shenhan Qian, Simon Giebenhain, Tim Walter, and Matthias Nie\ss{}ner.
\newblock Nersemble: Multi-view radiance field reconstruction of human heads.
\newblock \emph{ACM Trans. Graph.}, 42\penalty0 (4), 2023.

\bibitem[Li et~al.(2017)Li, Bolkart, Black, and Romero]{li2017learning}
Tianye Li, Timo Bolkart, Michael~J Black, and Javier Romero.
\newblock Learning a model of facial shape and expression from 4d scans.
\newblock 2017.

\bibitem[Lin et~al.(2020)Lin, Ryabtsev, Sengupta, Curless, Seitz, and Kemelmacher-Shlizerman]{BGMv2}
Shanchuan Lin, Andrey Ryabtsev, Soumyadip Sengupta, Brian Curless, Steve Seitz, and Ira Kemelmacher-Shlizerman.
\newblock Real-time high-resolution background matting.
\newblock \emph{arXiv}, pages arXiv--2012, 2020.

\bibitem[Livingstone and Russo(2018)]{livingstone2018ryerson}
Steven~R Livingstone and Frank~A Russo.
\newblock The ryerson audio-visual database of emotional speech and song (ravdess): A dynamic, multimodal set of facial and vocal expressions in north american english.
\newblock \emph{PloS one}, 13\penalty0 (5):\penalty0 e0196391, 2018.

\bibitem[Lombardi et~al.(2021)Lombardi, Simon, Schwartz, Zollhoefer, Sheikh, and Saragih]{lombardi2021mixture}
Stephen Lombardi, Tomas Simon, Gabriel Schwartz, Michael Zollhoefer, Yaser Sheikh, and Jason Saragih.
\newblock Mixture of volumetric primitives for efficient neural rendering.
\newblock \emph{ACM Trans. Graph.}, 40\penalty0 (4), 2021.

\bibitem[Martinez et~al.(2024)Martinez, Kim, Romero, Bagautdinov, Saito, Yu, Anderson, Zollhöfer, Wang, Bai, Li, Wei, Joshi, Borsos, Simon, Saragih, Theodosis, Greene, Josyula, Maeta, Jewett, Venshtain, Heilman, Chen, Fu, Elshaer, Du, Wu, Chen, Kang, Wu, Emad, Longay, Brewer, Shah, Booth, Koska, Haidle, Andromalos, Hsu, Dauer, Selednik, Godisart, Ardisson, Cipperly, Humberston, Farr, Hansen, Guo, Braun, Krenn, Wen, Evans, Fadeeva, Stewart, Schwartz, Gupta, Moon, Guo, Dong, Xu, Shiratori, Prada, Pires, Peng, Buffalini, Trimble, McPhail, Schoeller, and Sheikh]{martinez2024codec}
Julieta Martinez, Emily Kim, Javier Romero, Timur Bagautdinov, Shunsuke Saito, Shoou-I Yu, Stuart Anderson, Michael Zollhöfer, Te-Li Wang, Shaojie Bai, Chenghui Li, Shih-En Wei, Rohan Joshi, Wyatt Borsos, Tomas Simon, Jason Saragih, Paul Theodosis, Alexander Greene, Anjani Josyula, Silvio~Mano Maeta, Andrew~I. Jewett, Simon Venshtain, Christopher Heilman, Yueh-Tung Chen, Sidi Fu, Mohamed Ezzeldin~A. Elshaer, Tingfang Du, Longhua Wu, Shen-Chi Chen, Kai Kang, Michael Wu, Youssef Emad, Steven Longay, Ashley Brewer, Hitesh Shah, James Booth, Taylor Koska, Kayla Haidle, Matt Andromalos, Joanna Hsu, Thomas Dauer, Peter Selednik, Tim Godisart, Scott Ardisson, Matthew Cipperly, Ben Humberston, Lon Farr, Bob Hansen, Peihong Guo, Dave Braun, Steven Krenn, He Wen, Lucas Evans, Natalia Fadeeva, Matthew Stewart, Gabriel Schwartz, Divam Gupta, Gyeongsik Moon, Kaiwen Guo, Yuan Dong, Yichen Xu, Takaaki Shiratori, Fabian Prada, Bernardo~R. Pires, Bo Peng, Julia Buffalini, Autumn Trimble, Kevyn McPhail, Melissa Schoeller, and
  Yaser Sheikh.
\newblock {Codec Avatar Studio: Paired Human Captures for Complete, Driveable, and Generalizable Avatars}.
\newblock \emph{NeurIPS Track on Datasets and Benchmarks}, 2024.

\bibitem[Pan et~al.(2024)Pan, Zhuo, Piao, Luo, Cheng, Wang, Fan, Liu, Yang, Dai, Liu, Loy, Qian, Wu, Lin, and Lin]{pan2024renderme}
Dongwei Pan, Long Zhuo, Jingtan Piao, Huiwen Luo, Wei Cheng, Yuxin Wang, Siming Fan, Shengqi Liu, Lei Yang, Bo Dai, Ziwei Liu, Chen~Change Loy, Chen Qian, Wayne Wu, Dahua Lin, and Kwan-Yee Lin.
\newblock Renderme-360: A large digital asset library and benchmarks towards high-fidelity head avatars.
\newblock \emph{Advances in Neural Information Processing Systems}, 36, 2024.

\bibitem[Pang et~al.(2024)Pang, Zhu, Kortylewski, Theobalt, and Habermann]{pang2024ashanimatablegaussiansplats}
Haokai Pang, Heming Zhu, Adam Kortylewski, Christian Theobalt, and Marc Habermann.
\newblock Ash: Animatable gaussian splats for efficient and photoreal human rendering, 2024.

\bibitem[Prinzler et~al.(2024)Prinzler, Zakharov, Sklyarova, Kabadayi, and Thies]{prinzler2024jokerconditional3dhead}
Malte Prinzler, Egor Zakharov, Vanessa Sklyarova, Berna Kabadayi, and Justus Thies.
\newblock Joker: Conditional 3d head synthesis with extreme facial expressions, 2024.

\bibitem[Qian(2024)]{qian2024vhap}
Shenhan Qian.
\newblock Vhap: Versatile head alignment with adaptive appearance priors, 2024.

\bibitem[Ranjan et~al.(2018)Ranjan, Bolkart, Sanyal, and Black]{ranjan2018generating3dfacesusing}
Anurag Ranjan, Timo Bolkart, Soubhik Sanyal, and Michael~J. Black.
\newblock Generating 3d faces using convolutional mesh autoencoders, 2018.

\bibitem[Roich et~al.(2021)Roich, Mokady, Bermano, and Cohen-Or]{roich2021pivotaltuninglatentbasedediting}
Daniel Roich, Ron Mokady, Amit~H. Bermano, and Daniel Cohen-Or.
\newblock Pivotal tuning for latent-based editing of real images, 2021.

\bibitem[Saito et~al.(2023)Saito, Schwartz, Simon, Li, and Nam]{saito2023relightable}
Shunsuke Saito, Gabriel Schwartz, Tomas Simon, Junxuan Li, and Giljoo Nam.
\newblock Relightable gaussian codec avatars.
\newblock 2023.

\bibitem[Saito et~al.(2024)Saito, Schwartz, Simon, Li, and Nam]{saito2024relightablegaussiancodecavatars}
Shunsuke Saito, Gabriel Schwartz, Tomas Simon, Junxuan Li, and Giljoo Nam.
\newblock Relightable gaussian codec avatars, 2024.

\bibitem[Simonyan and Zisserman(2015)]{simonyan2015deepconvolutionalnetworkslargescale}
Karen Simonyan and Andrew Zisserman.
\newblock Very deep convolutional networks for large-scale image recognition, 2015.

\bibitem[Teotia et~al.(2024)Teotia, Kim, Garrido, Habermann, Elgharib, and Theobalt]{10.1145/3687927}
Kartik Teotia, Hyeongwoo Kim, Pablo Garrido, Marc Habermann, Mohamed Elgharib, and Christian Theobalt.
\newblock Gaussianheads: End-to-end learning of drivable gaussian head avatars from coarse-to-fine representations.
\newblock \emph{ACM Trans. Graph.}, 43\penalty0 (6), 2024.

\bibitem[Tewari et~al.(2017)Tewari, Zollhofer, Kim, Garrido, Bernard, Perez, and Theobalt]{Tewari_2017_ICCV}
Ayush Tewari, Michael Zollhofer, Hyeongwoo Kim, Pablo Garrido, Florian Bernard, Patrick Perez, and Christian Theobalt.
\newblock Mofa: Model-based deep convolutional face autoencoder for unsupervised monocular reconstruction.
\newblock In \emph{Proceedings of the IEEE International Conference on Computer Vision (ICCV) Workshops}, 2017.

\bibitem[Thies et~al.(2016)Thies, Zollhofer, Stamminger, Theobalt, and Niessner]{thies2016face2face}
Justus Thies, Michael Zollhofer, Marc Stamminger, Christian Theobalt, and Matthias Niessner.
\newblock Face2face: Real-time face capture and reenactment of rgb videos.
\newblock In \emph{Proceedings of the IEEE/CVF Conference on Computer Vision and Pattern Recognition (CVPR)}, pages 2387--2395, 2016.

\bibitem[Toussaint et~al.(2025)Toussaint, Thomas, and Franco]{toussaint2025probesdf}
Briac Toussaint, Diego Thomas, and Jean-S{\'e}bastien Franco.
\newblock Probesdf: Light field probes for neural surface reconstruction.
\newblock In \emph{Proceedings of the Computer Vision and Pattern Recognition Conference}, pages 11026--11035, 2025.

\bibitem[Tran and Liu(2018)]{tran2018nonlinear3dfacemorphable}
Luan Tran and Xiaoming Liu.
\newblock Nonlinear 3d face morphable model, 2018.

\bibitem[Tran et~al.(2019)Tran, Liu, and Liu]{tran2019highfidelitynonlinear3dface}
Luan Tran, Feng Liu, and Xiaoming Liu.
\newblock Towards high-fidelity nonlinear 3d face morphable model, 2019.

\bibitem[Vlasic et~al.(2005)Vlasic, Brand, Pfister, and Popovi\'{c}]{vlasic2005face}
Daniel Vlasic, Matthew Brand, Hanspeter Pfister, and Jovan Popovi\'{c}.
\newblock Face transfer with multilinear models.
\newblock \emph{ACM Trans. Graph.}, 24\penalty0 (3):\penalty0 426–433, 2005.

\bibitem[Wang et~al.(2022{\natexlab{a}})Wang, Wang, Sun, Kortylewski, and Yuille]{wang2022voge}
Angtian Wang, Peng Wang, Jian Sun, Adam Kortylewski, and Alan Yuille.
\newblock Voge: A differentiable volume renderer using gaussian ellipsoids for analysis-by-synthesis.
\newblock In \emph{The Eleventh International Conference on Learning Representations}, 2022{\natexlab{a}}.

\bibitem[Wang et~al.(2022{\natexlab{b}})Wang, Chandran, Zoss, Bradley, and Gotardo]{wang2022morf}
Daoye Wang, Prashanth Chandran, Gaspard Zoss, Derek Bradley, and Paulo Gotardo.
\newblock Morf: Morphable radiance fields for multiview neural head modeling.
\newblock In \emph{ACM SIGGRAPH 2022 Conference Proceedings}, pages 1--9, 2022{\natexlab{b}}.

\bibitem[Wang et~al.(2022{\natexlab{c}})Wang, Chen, Yu, Ma, Li, and Liu]{wang2022faceverse}
Lizhen Wang, Zhiyuan Chen, Tao Yu, Chenguang Ma, Liang Li, and Yebin Liu.
\newblock Faceverse: a fine-grained and detail-controllable 3d face morphable model from a hybrid dataset.
\newblock In \emph{Proceedings of the IEEE/CVF Conference on Computer Vision and Pattern Recognition (CVPR)}, 2022{\natexlab{c}}.

\bibitem[Xu et~al.(2023)Xu, Zhang, Wang, Zhao, Han, Guojun, and Liu]{xu2023latentavatar}
Yuelang Xu, Hongwen Zhang, Lizhen Wang, Xiaochen Zhao, Huang Han, Qi Guojun, and Yebin Liu.
\newblock Latentavatar: Learning latent expression code for expressive neural head avatar.
\newblock In \emph{ACM SIGGRAPH 2023 Conference Proceedings}, 2023.

\bibitem[Xu et~al.(2024)Xu, Wang, Zheng, Su, and Liu]{xu2024gphm}
Yuelang Xu, Lizhen Wang, Zerong Zheng, Zhaoqi Su, and Yebin Liu.
\newblock 3d gaussian parametric head model.
\newblock In \emph{Proceedings of the European Conference on Computer Vision (ECCV)}, 2024.

\bibitem[Yang et~al.(2020)Yang, Zhu, Wang, Huang, Shen, Yang, and Cao]{yang2020facescape}
Haotian Yang, Hao Zhu, Yanru Wang, Mingkai Huang, Qiu Shen, Ruigang Yang, and Xun Cao.
\newblock Facescape: a large-scale high quality 3d face dataset and detailed riggable 3d face prediction.
\newblock In \emph{Proceedings of the ieee/cvf conference on computer vision and pattern recognition}, pages 601--610, 2020.

\bibitem[Yenamandra et~al.(2021)Yenamandra, Tewari, Bernard, Seidel, Elgharib, Cremers, and Theobalt]{yenamandra2020i3dmm}
T Yenamandra, A Tewari, F Bernard, HP Seidel, M Elgharib, D Cremers, and C Theobalt.
\newblock i3dmm: Deep implicit 3d morphable model of human heads.
\newblock In \emph{Proceedings of the IEEE/CVF Conference on Computer Vision and Pattern Recognition (CVPR)}, 2021.

\bibitem[Zhang et~al.(2024)Zhang, Fang, Shrestha, Liang, Long, and Tan]{zhang2024rade}
Baowen Zhang, Chuan Fang, Rakesh Shrestha, Yixun Liang, Xiaoxiao Long, and Ping Tan.
\newblock Rade-gs: Rasterizing depth in gaussian splatting.
\newblock \emph{arXiv preprint arXiv:2406.01467}, 2024.

\bibitem[Zheng et~al.(2024{\natexlab{a}})Zheng, Gao, Fan, Liu, Laaksonen, Ouyang, and Sebe]{BiRefNet}
Peng Zheng, Dehong Gao, Deng-Ping Fan, Li Liu, Jorma Laaksonen, Wanli Ouyang, and Nicu Sebe.
\newblock Bilateral reference for high-resolution dichotomous image segmentation.
\newblock \emph{CAAI Artificial Intelligence Research}, 2024{\natexlab{a}}.

\bibitem[Zheng et~al.(2024{\natexlab{b}})Zheng, Wen, Li, Zhang, Su, Chang, Zhao, Lv, Zhang, Zhang, Wang, and Lan]{zheng2024headgap}
Xiaozheng Zheng, Chao Wen, Zhaohu Li, Weiyi Zhang, Zhuo Su, Xu Chang, Yang Zhao, Zheng Lv, Xiaoyuan Zhang, Yongjie Zhang, Guidong Wang, and Xu Lan.
\newblock Headgap: Few-shot 3d head avatar via generalizable gaussian priors.
\newblock \emph{arXiv preprint arXiv:2408.06019}, 2024{\natexlab{b}}.

\bibitem[Zheng et~al.(2025)Zheng, Wen, Li, Zhang, Su, Chang, Zhao, Lv, Zhang, Zhang, Wang, and Xu]{zheng2025headgapfewshot3dhead}
Xiaozheng Zheng, Chao Wen, Zhaohu Li, Weiyi Zhang, Zhuo Su, Xu Chang, Yang Zhao, Zheng Lv, Xiaoyuan Zhang, Yongjie Zhang, Guidong Wang, and Lan Xu.
\newblock Headgap: Few-shot 3d head avatar via generalizable gaussian priors, 2025.

\bibitem[Zhuang et~al.(2022)Zhuang, Zhu, Sun, and Cao]{zhuang2022mofanerf}
Yiyu Zhuang, Hao Zhu, Xusen Sun, and Xun Cao.
\newblock Mofanerf: Morphable facial neural radiance field.
\newblock In \emph{European conference on computer vision}, pages 268--285. Springer, 2022.

\end{thebibliography}
}
\clearpage
\setcounter{page}{1}
\maketitlesupplementary
\section{Dataset Details}
\label{sec:dataset_details}
\begin{figure}[h!]
    \centering    \includegraphics[width=\linewidth, height=\textheight, keepaspectratio]{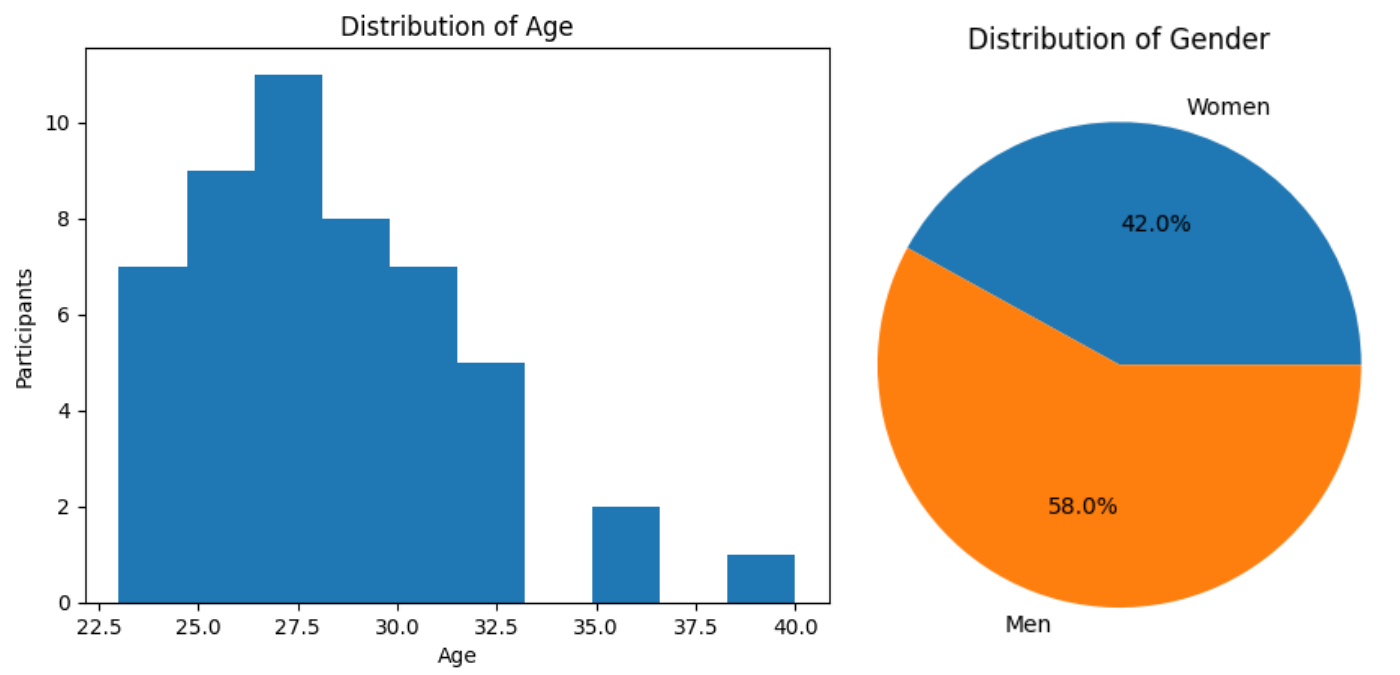}
    \caption{
        Statistics of the participants in our dataset.
    }
    \label{fig:age_gender}
\end{figure}

We utilise two datasets for training: EXPRESS-50 and RenderMe-360, each offering distinct advantages in terms of expression coverage, identity diversity, and multi-view supervision. Their combination enables robust learning of geometry, appearance, and expression disentanglement.
EXPRESS-50 is a multi-view dataset containing 50 subjects (29 male, 21 female), each performing 60 aligned expressions. Expressions are matched across all identities, enabling consistent expression conditioning during training. Subjects span ages 23–40 (mean: 28), with the age and gender distribution shown in Figure~\ref{fig:age_gender}.

\begin{figure}
    \centering
\includegraphics[width=\linewidth, keepaspectratio]{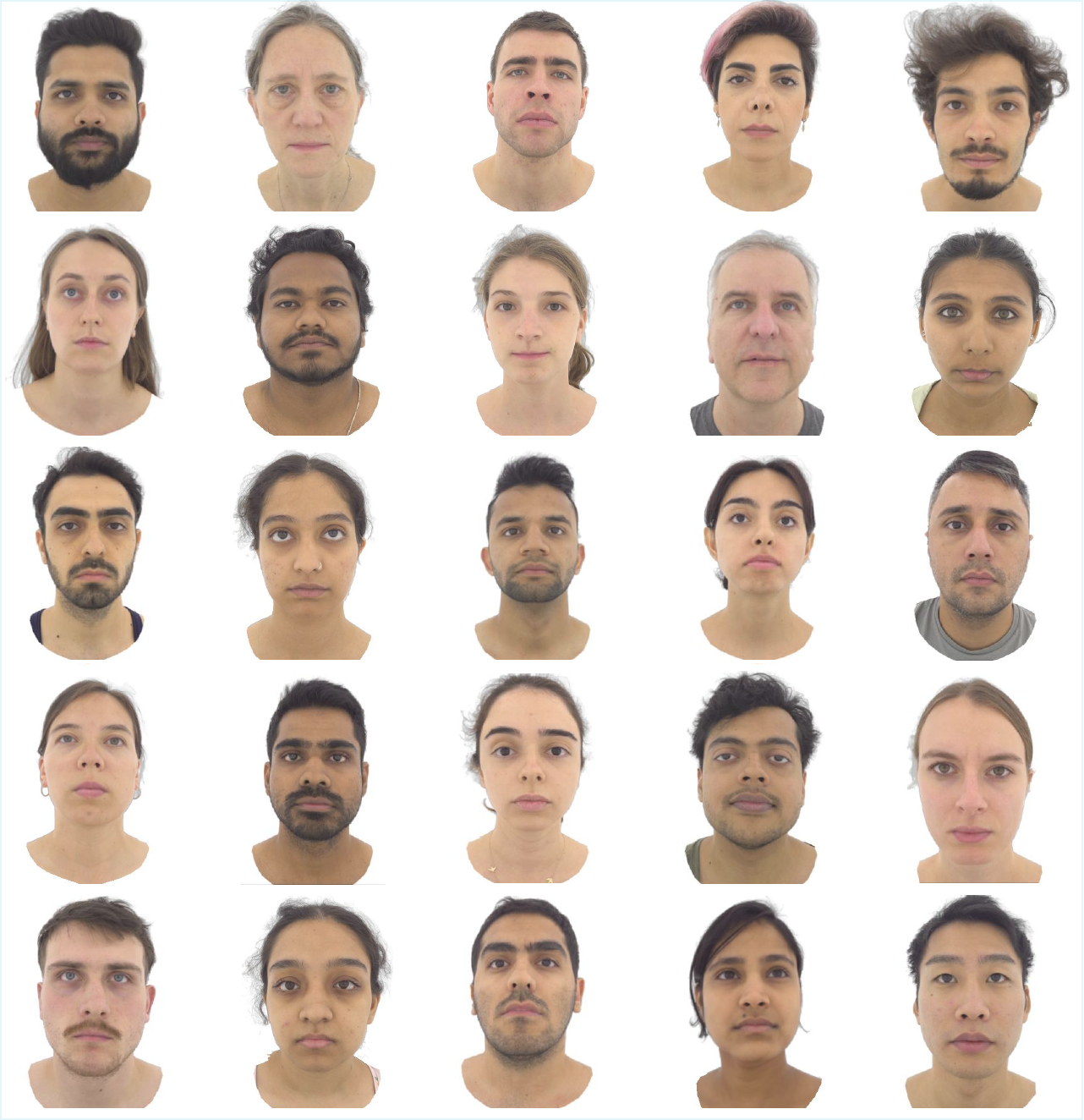}
    \caption{
Examples of 25 distinct identities from the EXPRESS-50 dataset, each aligned to the same neutral expression. This snapshot represents a subset of the 50 identities available in the dataset.}
    \label{fig:ours_id}
\end{figure}

\begin{figure*}
    \centering    \includegraphics[width=\textwidth, keepaspectratio]{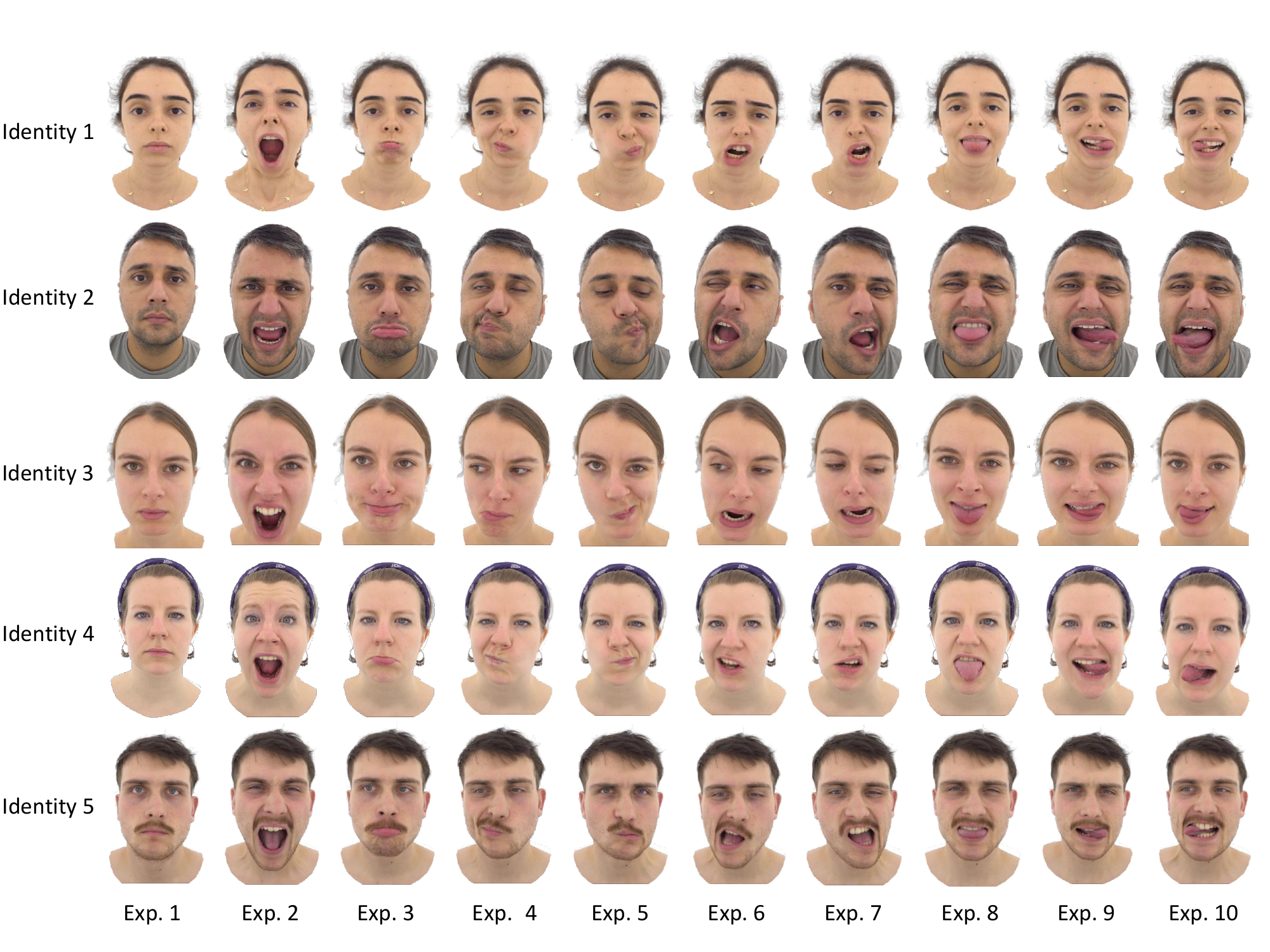}
    \caption{
        Diverse facial expressions from the EXPRESS-50 dataset are aligned consistently across all identities.
    }
    \label{fig:ours_exp}
\end{figure*}

Figure~\ref{fig:ours_id} shows different identities under a shared neutral expression, while Figure~\ref{fig:ours_exp} illustrates variation and consistency across 10 aligned expressions for five sampled subjects.
\begin{figure}[h!]
    \centering \includegraphics[width=\linewidth, keepaspectratio]{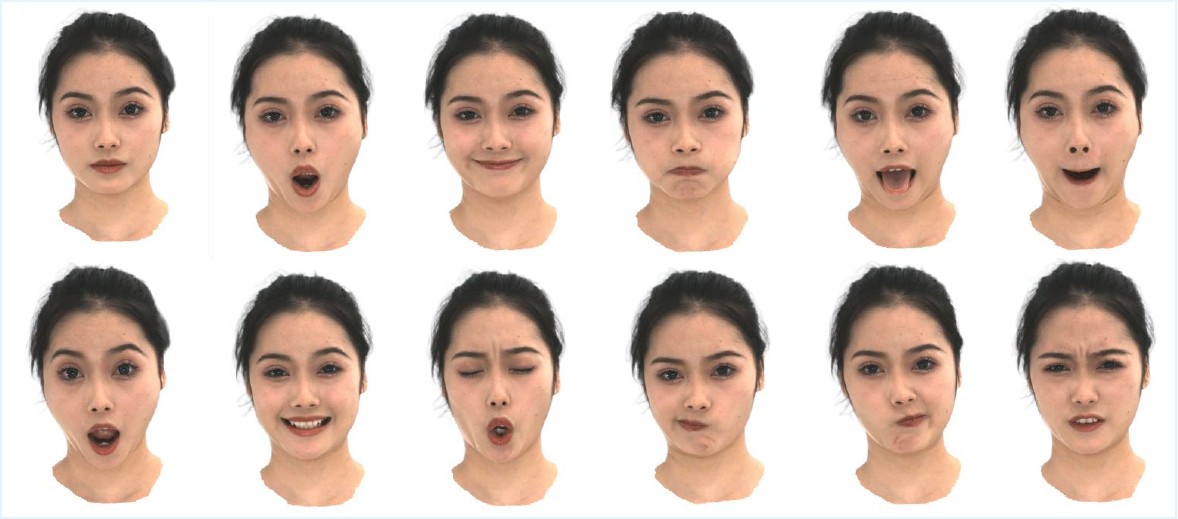}
    \caption{Examples of aligned expressions from the RenderMe360 dataset
    }
    \label{fig:renderme_exp}
\end{figure}

We select 280 identities from RenderMe-360 after filtering, each performing 12 semantically matched expressions under dense 360° multi-view capture. While it has fewer expressions per subject than EXPRESS-50, its rich identity and view coverage support generalisation across head poses and appearances. Example of aligned expressions from this dataset are shown in Figure~\ref{fig:renderme_exp}.

\section{Depth Supervision with ProbeSDF}
\begin{figure}[h!]
    \centering \includegraphics[width=\linewidth, keepaspectratio]{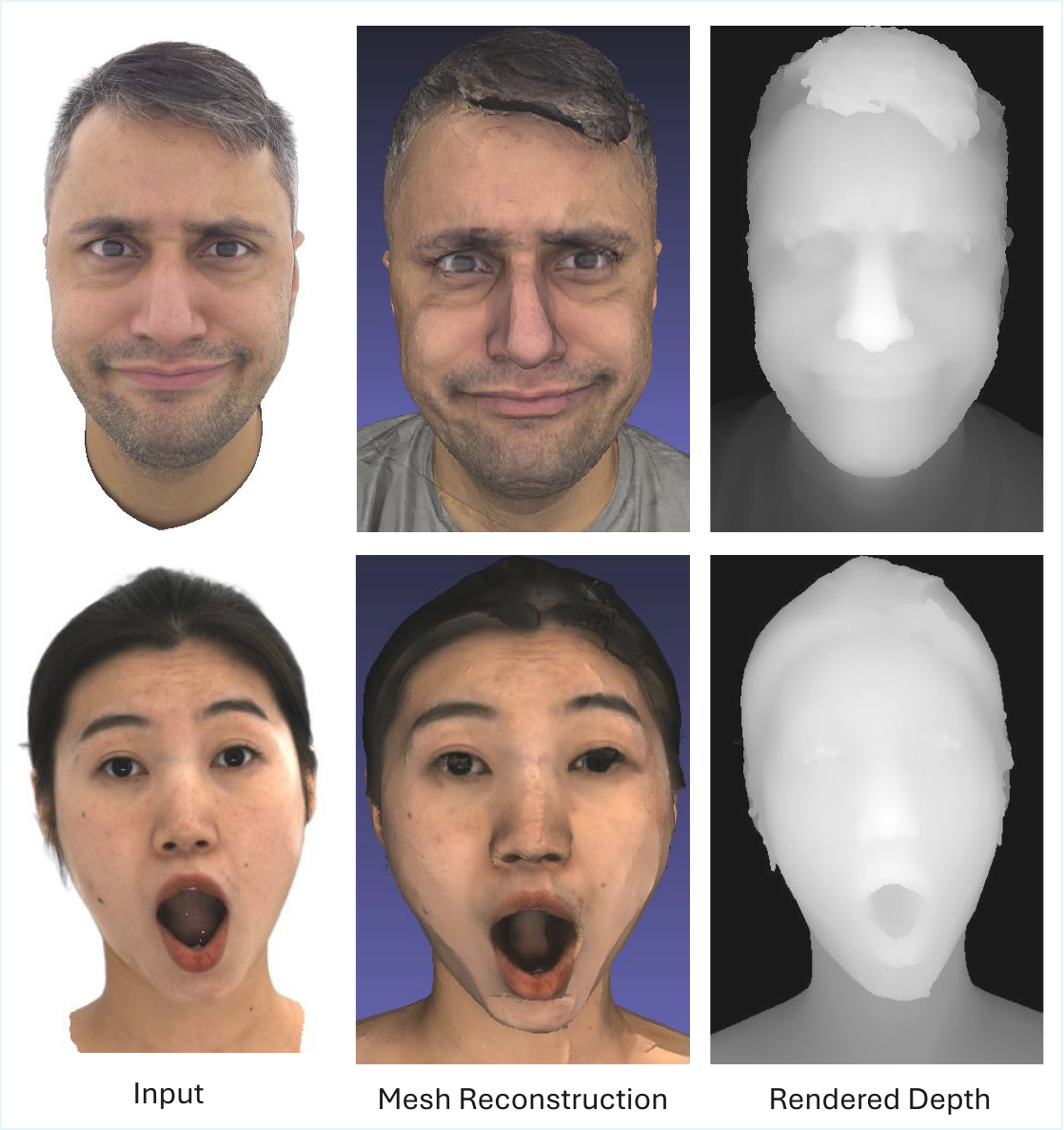}
    \caption{Examples of reconstructed depth from ProbeSDF.
    }
    \label{fig:depth_gt}
\end{figure}

To supervise geometry, we leverage ground-truth depth maps \( \mathbf{I}_{depth}^{gt} \) generated with ProbeSDF~\cite{toussaint2025probesdf}. These depth images are spatially aligned with the corresponding input RGB views, enabling us to directly measure consistency between the reconstructed geometry and the reference depth. As shown in Figure~\ref{fig:depth_gt}, ProbeSDF provides smooth and consistent depth for learning geometry.
\section{Mesh and UV Enhancement}
\label{sec:mesh_uv_enhancement}
\begin{figure}
\centering
\includegraphics[width=\linewidth]{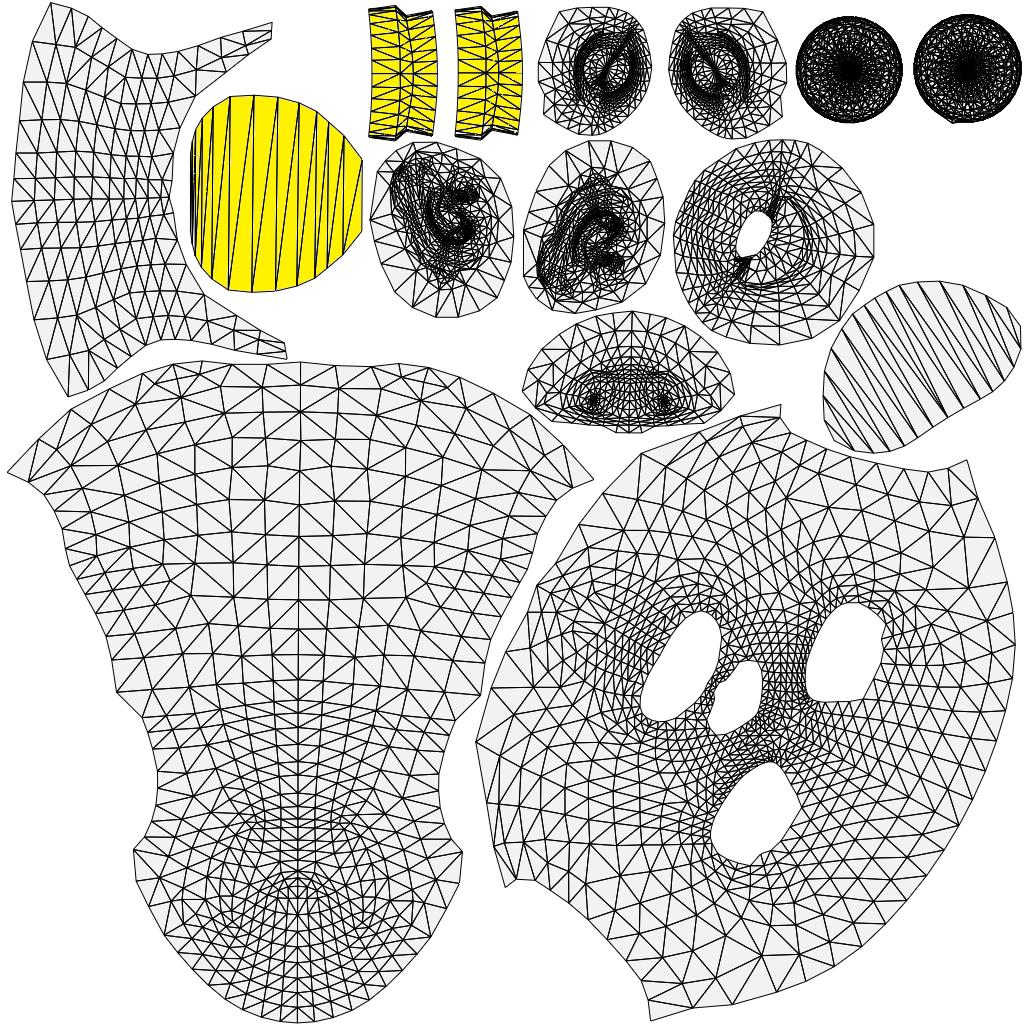}
\caption{\textbf{UV enhancement.} The mouth interior enhancement is highlighted in \textit{yellow}}
\label{fig:uv_enhancement}
\end{figure}

\begin{figure}
\centering
\includegraphics[width=\linewidth]{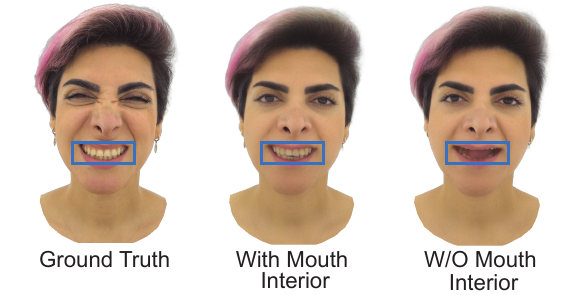}
\caption{Zeroing the expression residual $\mathbf{z}_{exp}$ exposes a shortcut: intra-oral appearance is entangled with expression, severely distorting the mouth.}
\label{fig:ablation_uv}
\end{figure}

Without explicit mouth–interior geometry, the model exploits a shortcut: Mouth interior geometry and appearance is implicitly encoded in the expression residual $\mathbf{z}_{exp}$, entangling expression and intra-oral appearance (see Figure~\ref{fig:ablation_uv}). Qualitatively, when we zero out the expression residual code $\mathbf{z}_{exp}$ the mouth interior becomes severely distorted, indicating that tooth and tongue detail is stored in the expression channel rather than in identity residuals $\mathbf{z}_{id}$. To improve geometric expressiveness and enable detailed modeling of the mouth interior, we extend the original FLAME~\cite{li2017learning} mesh topology by adding vertex sets for the upper and lower teeth, similar to VHAP~\cite{qian2024vhap}, along with a face for the inner mouth cavity. This modification modestly increases the vertex count while preserving FLAME’s semantic structure. To support Gaussian parameter prediction in these new regions, we augment the UV map to cover the extended topology. The resulting UV layout includes the mouth interior, allowing convolutional decoders to assign meaningful color, opacity, and feature values to mouth-interior Gaussians. Figure~\ref{fig:uv_enhancement} illustrates the enhanced UV layout.

\section{User study}
\label{sec:user_study}
We conducted a user study to qualitatively assess how well \sysname and Morphable Diffusion preserve (i) the reference person’s facial expression and (ii) identity. Each trial presented three images: a frontal reference (ground truth) on the left and two novel-view renderings from the two methods, labeled \textbf{A} and \textbf{B}. For every example, participants answered two questions: (1) which variant better preserves the expression, and (2) which variant better preserves the identity. Responses used five options: \emph{Strong preference for A/B}, \emph{Weak preference for A/B}, and \emph{Equally preferred} (tie). The assignment of \textbf{A} and \textbf{B} to the underlying methods was pre-specified and counterbalanced across examples to avoid label bias. We collected responses from \textbf{20} participants, each evaluating \textbf{15} examples (two judgments per example: expression and identity), yielding \textbf{600} total judgments. As summarized in Table~\ref{tab:user_study}, participants showed a clear overall preference for \sysname: \textbf{94.2\%} of judgments favored \sysname (sum of strong and weak preferences), \textbf{4.5\%} favored Morphable Diffusion, and \textbf{1.3\%} were ties.
\begin{table}[t]
\centering
\renewcommand{\arraystretch}{1}
\setlength{\tabcolsep}{3pt}
\begin{tabular}{c|cc|cc|c}
\toprule
\multirow{2}{*}{\textbf{Type}} & \multicolumn{2}{c|}{\textbf{Ours}} & \multicolumn{2}{c|}{\textbf{Morphable Diffusion}} & \multirow{2}{*}{\textbf{Tie}} \\
 & \textbf{++} & \textbf{+} & \textbf{++} & \textbf{+} &  \\
\midrule
Expression & 86.7\% & 6.0\% & 3.7\% & 1.0\% & 2.7\% \\
Identity   & 89.3\% & 6.3\% & 2.7\% & 1.7\% & 0.0\% \\
\midrule
Overall    & 88.0\% & 6.2\% & 3.2\% & 1.3\% & 1.3\% \\
\bottomrule
\end{tabular}
\caption{User preference distribution comparing \textbf{Ours} and \textbf{Morphable Diffusion} on expression and identity preservation. Percentages are pooled over all judgments; ``++'' denotes strong preference and ``+'' denotes weak preference. Overall: \textbf{Ours} 94.2\%, \textbf{Morphable Diffusion} 4.5\%, Tie 1.3\%.}
\label{tab:user_study}
\end{table}

\section{Locality regularisation for inverse rendering}
\label{sec:locality}
\begin{figure}[h!]
\centering
\includegraphics[width=\linewidth]{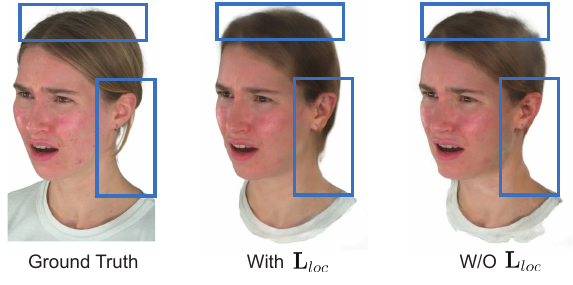}
\caption{\textbf{Locality loss preserves the prior and improves novel views.} Qualitative ablation comparing \emph{With locality loss} vs. \emph{W/O locality loss}. The locality loss preserves the prior and yields sharper detail and fewer artifacts in novel views. PSNR (dB): \emph{With} 28.43 vs. \emph{W/O} 27.75 (\(+0.68\)). \textit{Please zoom in for details.}}
\label{fig:ablation_loc}
\end{figure}

\paragraph{Locality regularization.}
During the second-stage refinement, we regularize the model to remain close to the pretrained solution for interpolations between the fine-tuned subject and dataset identities. 
Specifically, we sample a dataset identity $r$ with known parameters 
$(\mathbf{z}_{id}^{r}, \mathbf{z}_{exp}^{r}, \alpha_{id}^{r}, \alpha_{exp}^{r}, \theta_{jaw}^{r}, \theta_{neck}^{r})$ 
and consider the fine-tuned subject 
$(\mathbf{z}_{id}^{*}, \mathbf{z}_{exp}^{*}, \alpha_{id}^{*}, \alpha_{exp}^{*}, \theta_{jaw}^{*}, \theta_{neck}^{*})$. 
We construct interpolated parameters at a fixed $t=0.3$:
\begin{align*}
\tilde{\mathbf{z}}_{id} &= (1-t)\,\mathbf{z}_{id}^{r} + t\,\mathbf{z}_{id}^{*}, \\
\tilde{\mathbf{z}}_{exp} &= (1-t)\,\mathbf{z}_{exp}^{r} + t\,\mathbf{z}_{exp}^{*}, \\
\tilde{\alpha}_{id} &= (1-t)\,\alpha_{id}^{r} + t\,\alpha_{id}^{*}, \\
\tilde{\alpha}_{exp} &= (1-t)\,\alpha_{exp}^{r} + t\,\alpha_{exp}^{*}, \\
\tilde{\theta}_{jaw} &= \mathrm{SLERP}(\theta_{jaw}^{r},\, \theta_{jaw}^{*},\, t), \\
\tilde{\theta}_{neck} &= \mathrm{SLERP}(\theta_{neck}^{r},\, \theta_{neck}^{*},\, t).
\end{align*}
We render two images using the same interpolated parameters: 
$I_{frozen}$ with the pretrained frozen model, and $I_{tuned}$ with the fine-tuned model. 
Our \emph{locality loss} is then defined as:
\[
\mathbf{L}_{loc} = \mathbf{L}_{rec}(I_{tuned},\, I_{frozen}),
\]
where $\mathbf{L}_{rec}$ is the image-space reconstruction loss defined in Sec.~\ref{sec:training}.
This encourages the refined model to preserve the behaviour of the pretrained prior (Figure ~\ref{fig:ablation_loc}) along interpolation paths between known dataset identities and the personalized subject, ensuring that refinements remain localized.

\section{Additional ablations.}
\label{sec:additional_ablation}
\begin{figure}[h!]
\centering
\includegraphics[width=\linewidth]{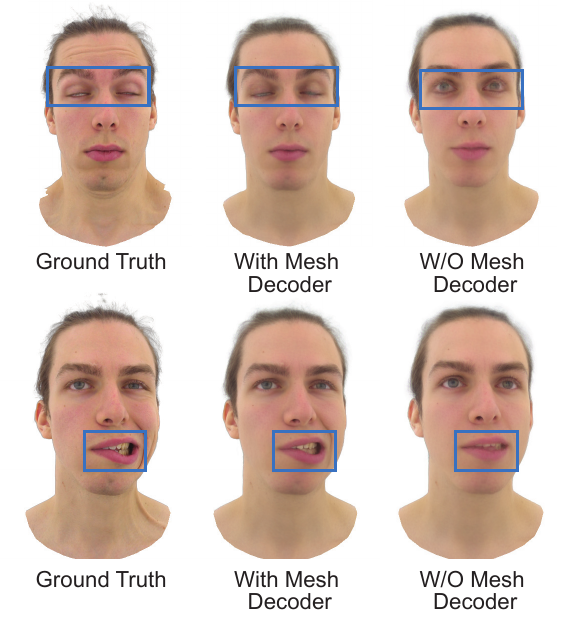}
\caption{\textbf{No mesh decoder harms expressivity and realism.} Qualitative ablation comparing \emph{With mesh decoder} vs. \emph{W/O mesh decoder}. Disabling $\Phi_{\mathrm{mesh}}$ and learning only residual Gaussian offsets weakens mouth and cheek articulation and degrades photorealism. PSNR (dB): \emph{W/O mesh decoder} 32.34 vs. \emph{With mesh decoder} 34.52 (\(+2.18\)). \textit{Please zoom in for details.}}
\label{fig:ablation_vertex}
\end{figure}

\paragraph{No Mesh Decoder.}
Disabling $\Phi_{\mathrm{mesh}}$ and learning only fine residual offsets for the Gaussian primitives reduces mouth and cheek articulation, which yields poorer facial expressivity and reduced photorealism, as shown in Figure~\ref{fig:ablation_vertex}.

\begin{figure}
\centering
\includegraphics[width=\linewidth]{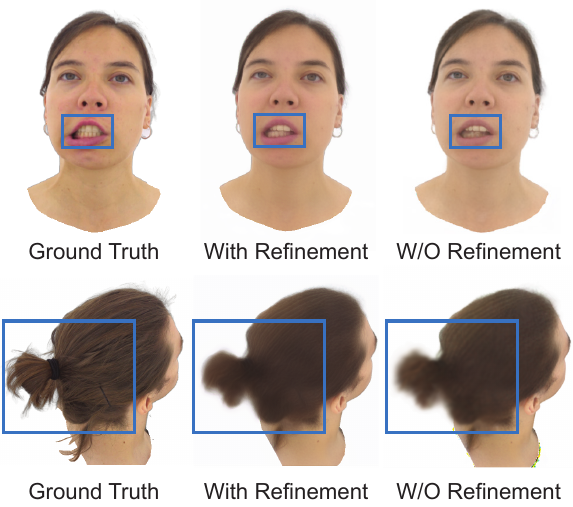}
\caption{\textbf{Screen-space refinement improves fidelity.} Qualitative ablation comparing \emph{With} vs. \emph{W/O} $\Psi_{\mathrm{ref}}$. Eliminating $\Psi_{\mathrm{ref}}$ degrades hair texture, removes mouth details, and reduces overall sharpness and fidelity—even after 3D-aware rasterization. PSNR (dB): \emph{W/O} 30.83 vs. \emph{With} 31.60 (\(+0.77\)). \textit{Please zoom in for details.}}
\label{fig:ablation_refine}
\end{figure}

\paragraph{No Refinement Network.}
Eliminating $\Psi_{\mathrm{ref}}$ leads to degraded hair texture, loss of mouth details, and reduced overall sharpness and fidelity, as shown in Figure~\ref{fig:ablation_refine}. This highlights the importance of screen-space correction, even after  rasterization.

\section{Implementation details.}
The model is trained for 250{,}000 iterations with a batch size of 1 on four NVIDIA A100 GPUs. We use the Adam~\cite{kingma2017adammethodstochasticoptimization} optimizer with a learning rate of $1 \times 10^{-4}$ for all learnable parameters.

\section{Additional Results.}
\label{sec:additional_results}
\begin{figure}[h!]
\centering
\includegraphics[width=\linewidth]{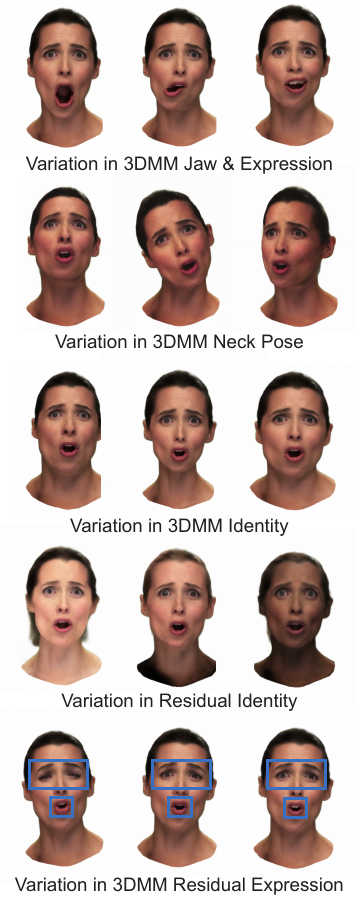}
\caption{Distangled control of \sysname parameters.}
\label{fig:disentangle_control}
\end{figure}

\paragraph{Disentangled control.}
We demonstrate disentangled control over 3DMM and residual parameters; see Figure~\ref{fig:disentangle_control}.
\section{Limitations and Future Work}
While significantly advancing the state of the art, our model is not without limitations.
One notable limitation of the method lies in its difficulty in handling out-of-distribution subjects, such as individuals with long hair or unconventional facial features, which may deviate significantly from the training data. Additionally, variations in the lighting environment can challenge the model's robustness, potentially leading to artefacts or inaccuracies in rendering. These issues underscore the need to enhance the generalizability of the approach. A promising direction to address this limitation involves annotating a more diverse set of identities with a wide range of expressions and lighting variations, ensuring that the model can better accommodate subjects with varying appearances and environmental conditions, thereby enhancing the method's ability to generalise to more challenging real-world scenarios.

\end{document}